\documentclass{emulateapj}
 \newcommand{\Mdot}{\dot{M}}
\newcommand{\Msun}{M_{\odot}} 

\newcommand{\csdisk}{c_{s,d}}

\newcommand{\pomega}{\varpi}
\newcommand{\mach}{\mathcal{M}}
\newcommand{\dt}{\frac{\partial}{\partial t}}
\newcommand{\vecx}{\mathbf{x}}
\newcommand{\vecv}{\mathbf{v}}
\newcommand{\vecp}{\mathbf{p}}

\begin{document}
\title{On the Role of Disks in the Formation of Stellar Systems: a numerical parameter study of rapid accretion. }
\author{Kaitlin M. Kratter, Christopher D. Matzner}
\affil{Department of Astronomy and Astrophysics,  50 St. George Street, University of Toronto, Toronto, Ontario, M5R 3H4, Canada}
\author{Mark R. Krumholz}
\affil{Department of Astronomy,  201 Interdisciplinary Sciences Building, University of California, Santa Cruz, CA 95064, USA}
\author{Richard I.  Klein}
\affil{Department of Astronomy, University of California, Berkeley 
601 Campbell Hall, Berkeley, CA 94720-3411,USA \& Lawrence Livermore National Laboratory, AX Division 
7000 East Avenue, Livermore, CA 94550, USA }


\begin{abstract} 
We study rapidly accreting, gravitationally unstable disks with a series of global, three dimensional,  numerical experiments using the code ORION.  In this paper we conduct a numerical parameter study focused on protostellar disks, and show that one can predict disk behavior and the multiplicity of the accreting star system as a function of two dimensionless parameters which compare the disk's accretion rate to its sound speed and orbital period.  Although gravitational instabilities become strong,  we find that fragmentation into binary or multiple systems occurs only when material falls in several times more rapidly than the canonical isothermal limit. The disk-to-star accretion rate is proportional to the infall rate, and governed by gravitational torques generated by low-$m$ spiral modes. We also confirm the existence of a maximum stable disk mass: disks that exceed $\sim 50\%$ of the total system mass are subject to fragmentation and the subsequent formation of binary companions. 
\end{abstract}

\section{Introduction}
Accretion disks are ubiquitous in astrophysical systems, feeding objects ranging in size from planets to black holes.  Rapid disk accretion requires rapid angular momentum transport, but only a few transport mechanisms are known: self-gravity, turbulence from the magnetorotational instability, driven hydrodynamical turbulence, and (in rare circumstances) magneto-centrifugal winds.   Our focus is on the dynamics of disks around young, rapidly accretings protostars, for which self-gravity is the key ingredient  \citep{1987MNRAS.225..607L, Gam2001, KMK08}.  Gravitational instability (hereafter GI) is important whenever disks are cold enough or massive enough to trigger it, as they typically are during the early phases of star formation. 
GI plays a strong role in AGN disks as well, and possibly in other contexts where disks are cold and accretion is fast.

The role of GI in angular momentum transport is complicated by the fact that it can lead to runaway collapse.  Indeed, this makes GI an attractive mechanism for the formation of stars with companions
\citep[binaries, brown dwarfs, even planets in some circumstances:][]{1994MNRAS.269L..45B,1994MNRAS.271..999B,2006astro.ph..2367W}.  \citealt{KM06} (2006, hereafter KM06) and \citealt{KMK08} (2008, hereafter KMK08)
 found that disk fragmentation and binary formation are increasingly likely as one considers more and more massive stars, whereas disks in low-mass star formation are relatively stable \citep{ML2005}. 
The increased frequency of giant planet formation around A stars relative to F and G stars, and its lower sensitivity to the stellar metallicity  \citep{2007ApJ...665..785J}, may also be related.

During the earliest phases of star formation, protostellar disks are deeply embedded within their natal clouds.  
Observing this stage has been difficult because the source is only visible at wavelengths where resolution is poor.   But because it sets the initial conditions for stellar and planetary systems, an understanding of this phase is critical.  Models of rapidly accreting disks, like those presented here, will be useful for the interpretation of observations of future facilities such as ALMA and the EVLA.  

Although semi-analytical and low-dimensional studies can illuminate trends and provide useful approximate results, disk fragmentation is inherently a nonlinear and multidimensional process.   
For this reason we have embarked on
a survey of global, three-dimensional, numerical experiments to examine the role of GI as the mediator of the accretion rate in self-gravitating disks, and as a mechanism for creating 
disk-born companions.     

Any such project faces a central difficulty: the GI is famously sensitive to the disk's thermodynamics \citep{2000ApJ...528..325B,Gam2001,2003ApJ...590.1060P,Rice05,2005ApJ...619.1098M,Krum2006b,2008ApJ...673.1138C}.   While it is possible and valuable to incorporate detailed heating and cooling into numerical simulations as has been explored by the authors above, there is a cost: simulations with these important physical processes cannot be scaled to represent a whole range of physical environments, whereas those without them can.   We choose to separate the dynamical problem from the thermal one.  We exclude thermal physics from our simulations entirely, while scanning a thermal parameter in our survey.   
By this means we reduce the physical problem to two dimensionless parameters: one for the disk's temperature, another for its rotation period -- both in units determined by its mass accretion rate. We hold these fixed in each simulation by choosing well-controlled initial conditions corresponding to self-similar core collapse.  This parameterization is a central aspect of our work: it forms the basis for our numerical survey; it allows us to treat astrophysically relevant disks, including fragmentation and the formation of binary companions, while also maintaining generality; and it distinguishes our work from previous numerical studies of core collapse, disk formation, GI, and fragmentation.  

This paper, the first in a series, focuses on the broad conclusions we can draw from our parameter space study; subsequent papers will discuss the detailed behavior of multiple systems, three dimensional effects such as turbulence, and vertical flows, and non-linear GI mode coupling. We begin here by introducing our dimensionless parameters in \S \ref{survey}. We describe the initial conditions and the numerical code used in \S \ref{code}. In \S \ref{scalings} we derive analytic predictions for the behavior of disks as a function of our parameters. We describe the main results from our numerical experiments in \S \ref{results}, with more detailed analysis in \S \ref{caveats}. We compare them in detail to other numerical and analytic models of star formation in \S \ref{prevwork}. 

\section{A New Parameter Space for Accretion} \label{survey}

We consider the gravitational collapse of a rotating, quasi-spherical gas core onto a central pointlike object, mediated by
a disk. In the idealized picture we will explore in this paper, the disk and the mass flows into and out of it can be characterized by a
few simple parameters. At any given time, the central point mass (or
masses, in cases where fragmentation occurs) has mass $M_*$, the disk
has mass $M_d$, and the combined mass of the two is $M_{*d}$. The disk
is characterized by a constant sound speed $c_{s,d}$. Material from
the core falls onto the disk with a mass accretion rate $\dot{M}_{\rm
in}$, and this material carries mean specific angular momentum
$\langle j \rangle_{\rm in}$, and as a result it circularizes and goes
into Keplerian rotation at some radius $R_{k, \rm in}$; the angular
velocity of the orbit is $\Omega_{k, \rm in}$. In general in what
follows, we refer to quantities associated with the central object
with subscript *, quantities associated with the disk with a subscript
d, quantities associated with infall with subscript in. Angle brackets
indicate mass-weighted averages over the disk (with subscript d) or over infalling mass
(with subscript in).

Our simple decomposition of the problem is motivated by the work of \cite{Gam2001},  \cite{ML2005}, and KMK08. We characterize our numerical experiments using two dimensionless parameters which are well-adapted to systems undergoing rapid accretion.  We encapsulate the complicated physics of heating and cooling through the thermal parameter 
  \begin{equation}\label{xi}
 \xi = \frac{\Mdot_{\rm in} G}{\csdisk^3},
  \end{equation}
which relates the infall mass accretion rate $\Mdot_{\rm in}$ to the characteristic sound speed $\csdisk$ of the disk material.   Our parameter $\xi$ is also related to the physics of core collapse leading to star formation.  If the initial core is characterized by a signal speed $c_{{\rm eff},c}$  then $\Mdot_{\rm in}\sim c_{{\rm eff},c}^3/G$, implying $\xi\sim c_{{\rm eff},c}^3/\csdisk^3$ -- although there can be large variations around this value \citep{1972MNRAS.156..437L,1993ApJ...416..303F}.  

The second, rotational parameter
\begin{equation}\label{gamma}
\Gamma = \frac{\Mdot_{\rm in}}{M_{*d} \Omega_{k,{\rm in}} }  = \frac{\Mdot_{\rm in} \left<j\right>_{\rm in}^3 }{G^2 M_{*d}^3},
\end{equation}
compares the system growth rate or accretion timescale, $\Mdot_{\rm in}/M_{*d}$ to the orbital timscale of infalling gas.  Unlike $\xi$, $\Gamma$ is independent of disk heating and cooling, depending only on the core structure and velocity field. In general, $\Gamma$  compares the relative strength of rotation and gravity in the core.  Systems with a large value of $\Gamma$ (e.g. accretion-induced collapse of a white dwarf) gain a significant amount of mass in each orbit, and tend to be surrounded by thick, massive accretion disks, while those with very low $\Gamma$ (e.g. active galactic nuclei) grow over many disk lifetimes, and tend to harbor thin disks with little mass relative to the central object.  We consider characteristic values for our parameters in \S \ref{thepars}, and their evolution in the isothermal collapse of a rigidly rotating Bonnor-Ebert sphere in \S \ref{bonnorebert}. 

In addition to being physically motivated, our parameters are practical from a numerical and observational perspective. The thermal parameter $\xi$ is straightforward to calculate in other simulations, theoretical models, and observed disks. Accretion rates are routinely estimated from measures of infall velocities in cores \citep[although some uncertainties persist]{cesaroni07a}. Disk temperatures can also be measured using infrared and submillimeter disk detections. By contrast, measuring classic dimensionless disk parameters such as Toomre's $Q = c_s \Omega/ (\pi G \Sigma)$ \citep{Toom1964} can be difficult. In observed disks, estimating $Q$ is challenging due to the current resolution and sensitivity of even the best instruments.  While constraining disk temperatures is possible, measuring accurate (within a factor of $\sim 3$) surface densities are not \citep{2008ApJ...683..304E}. The rotation parameter $\Gamma$ is similarly practical: one need not know the density distribution of the initial core, nor the radial and angular distribution of the velocity profile -- a mean value for $j$ and an estimate of the core mass is sufficient to make an estimate for the disk size, and thus $\Gamma$. 

To study the evolution of systems as they accrete, we hold $\xi$ and $\Gamma$ fixed for each experiment via the self-similar collapse of a rotating, isothermal sphere (\S \ref{selfsim}).     This strategy allows us to map directly between the input parameters, and relevant properties of the system.  Specifically, we expect dimensionless properties like the disk-to-star mass ratio, Toomre parameter, stellar multiplicity, etc., to fluctuate around well-defined mean values (see \S \ref{selfsim_idea}).

We aim to use our parameters $\xi$ and $\Gamma$ to: (a)  explore the parameter space relevant to a range of star formation scenarios; (b) better understand the disk parameters, both locally and globally, which dictate the disk accretion rate and fragmentation properties; (c) make predictions for small scale disk behavior based on larger scale, observable quantities; and (d) allow the results of more complicated and computationally expensive simulations to be extended into other regimes.

\subsection{Characteristic values of the accretion parameters} \label{thepars}

We base our estimates of $\Gamma$ and $\xi$ on observations of core rotation in low-mass  and massive star-forming regions \citep{1992ApJ...396..631M, 1993ApJ...406..528G,1999ApJ...511..208W}, as well as the analytical estimates of core rotation and disk temperature in \cite{ML2005},  \cite{Krum2006b}, KM06, and KMK08.   Using simple models of core collapse in which angular momentum is conserved in the collapse process and part of the matter is cast away by protostellar outflows \citep{MM2000}, we find that both $\xi$ and $\Gamma$ are higher in massive star formation than in low-mass star formation.  In our models, the characteristic value of $\Gamma$ rises from $\sim 0.001 - 0.03$ as one considers increasingly massive cores for which turbulence is a larger fraction of the initial support.  

The value of $\xi$ is more complicated, as it reflects the disk's thermal state as well as infalling accretion rate, but the models of KMK08 and  \cite{Krum2006b} indicate that its characteristic value increases from $\lesssim 1$ to $\sim 10$ as one considers higher and higher mass cores -- although the specific epoch in the core's accretion history is also important.  In the case of massive stars, such rapid accretion has been observed as in \cite{2006Natur.443..427B} and \cite{2008arXiv0812.1789B}. Numerical simulations also find rapid accretion rates from cores to disks.  Simulations such as those of \cite{BanPud07} report $\xi  \sim 10$  at early times in both magnetized and non-magnetized models. We note that $\Gamma$ has significant fluctuations from core to core when turbulence is the source of rotation, and both $\xi$ and $\Gamma$ are affected by variations of the core accretion rate around its characteristic value \citep{1993ApJ...416..303F}.

  A major goal of this work is to probe the evolution of disks with $\xi \geq 1$, as mass accretion at this rate cannot be accommodated by the \cite{SS1973} model with $\alpha<1$.  Values of $\alpha$ exceeding unity imply very strong GI, and possibly fragmentation. 

\section{Numerical Methodology}\label{code}
\subsection{Numerical Code}\label{code_details}
We use the code ORION to conduct our numerical experiments \citep{Truelove98,Klein99,  2002PhDT.........5F}. ORION is a parallel 
adaptive mesh refinement (AMR), multi-fluid, radiation-hydrodynamics code with self-gravity and lagrangian sink particles (Krumholz et al. 2004).  Radiation transport and multi-fluids 
are not used in the present study. The gravito-hydrodynamic equations are solved using a conservative, Godunov scheme, which is second order accurate in both space and time.  The gravito-hydrodynamic equations are:
\begin{eqnarray}
\dt \rho & = & -\nabla \cdot (\rho\vecv) - \sum_i \dot{M}_i W(\vecx-\vecx_i)
\label{masscons}
\\
\dt (\rho\vecv) & = & -\nabla \cdot (\rho\vecv\vecv) - \nabla P - \rho\nabla \phi \\ \nonumber
& -& \sum_i  \dot{\vecp}_i W(\vecx-\vecx_i) \\ \nonumber
\label{gasmom}
\\
\dt (\rho e) & = & -\nabla \cdot [(\rho e + P)\vecv] + \rho \vecv \cdot \nabla \phi  \\ \nonumber
&-&\sum_i \dot{\mathcal{E}}_i W(\vecx-\vecx_i)
\label{gasen}
\end{eqnarray}

Equations (\ref{masscons})-(\ref{gasen}) are the equations of mass, momentum and energy conservation respectively. In the equations above, $\Mdot_i$, $\dot{\vecp}_i$, and $\dot{\mathcal{E}}_i$ describe the rate at which mass and momentum are transfered from the gas onto the $i$th lagrangian sink particles. Summations in these equations are over all sink particles present in the calculation. $W(\vecx)$ is a weighting function that defines the spatial region over which the particles interact with gas. The corresponding evolution equations for sink particles are

\begin{eqnarray}
\frac{d}{dt} M &=& \Mdot_i \\
\frac{d}{dt} \vecx_i & = & \frac{\vecp_i}{M_i}\\
\frac{d}{dt} \vecp_i &= &-M_i \nabla \phi +\dot{\vecp_i}.
\end{eqnarray}
These equations describe the motion of the point particles under the influence of gravity while accreting mass and momentum from the surrounding gas.

The Poisson equation is solved by multilevel elliptic solvers via the multigrid method.  The potential $\phi$ is given by the Poisson equation
\begin{equation}
\label{poisson}
\nabla^2 \phi = 4 \pi G \left[\rho + \sum_i M_i \delta(\vecx-\vecx_i)\right],
\end{equation}
and the gas pressure $P$ is given by
\begin{equation}
\label{pres}
P = \frac{\rho k_{\rm B} T_{\rm g}}{\mu} = (\gamma - 1) \rho \left(e - \frac{1}{2} v^2\right),
\end{equation}
where $T_{\rm g}$ is the gas temperature, $\mu$ is the mean particle mass, and $\gamma$ is the ratio of specific heats in the gas. We adopt $\mu=2.33 m_{\rm H}$, which is appropriate for standard cosmic abundances of a gas of molecular hydrogen and helium.

We use the sink particle implementation described in \cite{Krumholz04} to replace cells which become too dense to resolve. Sink particle creation and AMR grid refinement are based on the Truelove criterion \citep{Truelove97} which defines the maximum density that can be well resolved in a grid code as:
\begin{equation}\label{truelove}
\rho < \rho_j = \frac{N_J^2\pi c_s^2 } {G (\Delta x^l)^2},
\end{equation}
where, $N_J$ is the Jeans number, here set to $0.125$ for refinement, and $0.25$ for sink creation, and $\Delta x^l$ is the cell size on level $l$. When a cell violates the Jeans criterion, the local region is refined to the next highest grid level. If the violation occurs on the maximum level specified in the simulation, a sink particle is formed. Setting $N_J$ to 0.125 is also consistent with the resolution criterion in \cite{Nelson2006}. Sink particles within 4 cells of each other are merged in order to suppress unphysical n-body interactions due to limited resolution. At low resolution, unphysical sink particle formation and merging, can cause rapid advection of sink particles inwards onto the central star, generating spurious accretion. Moreover, because an isothermal, rotating gas filament will collapse infinitely to a line \citep{Truelove97}, an entire spiral arm can fragment and be merged into a single sink particle. To alleviate this problem, we implement a small barotropic switch in the gas equation of state such that
\begin{eqnarray}\label{baroswitch}
\gamma &=& 1.0001,~ \rho < \rho_{J^{s}}/4 \\
\gamma &= &1.28,~ \rho_{J^s}/4 < \rho < \rho_{J^s},
\end{eqnarray}
where the $J^s$ subscript indicates the Jean's criterion used for sink formation.
With this prescription, gas is almost exactly isothermal until fragmentation is imminent,  at which point it stiffens somewhat. 
This modest stiffening helps turn linear filaments into resolved spheres just prior to collapse and provides separation between newborn sink particles. 
The primary effect of this stiffening is to increase the resolution of the most unstable wavelength in a given simulation, at the expense of some dynamical range. We describe the influence of this stiffening on our results in \S \ref{thermo}, where we conduct some experiments in which it is turned off. 

 As described via equations (\ref{masscons})-(\ref{gasen}), sink particles both accrete from and interact with the gas and each other via gravity. Accretion rates are computed using a modified Bondi-Hoyle formula which prevents gas which is not gravitationally bound to the particles from accreting.  See \cite{Krumholz04} and \cite{Offner08} for a detailed study of the effects of sink particle parameters. Note that we also use a secondary, spatial criterion for AMR refinement based on an analytic prediction for the disk size as a function of time (see \S \ref{domain}).

 \subsection{Initial conditions}\label{selfsim}
We initialize each run with an isothermal core: 
\begin{eqnarray}\label{iso_shu}
\rho (r) = \frac {A c_{s,{\rm core}}^2}{4 \pi G r^2}. 
\end{eqnarray}
There is a small amount of rotational motion in our initial conditions, but no radial motion.  A core with this profile is out of virial balance when $A>2$, and accretes at a rate 
\begin{equation}\label{MdotCore} 
\dot{M} = {c_{s, {\rm core}}^3\over G} \times 
\left\{  
\begin{array}{lc}
    0.975, ~& (A=2) \\   
(2A)^{3/2} /\pi. &(A \gg 2)
\end{array}
\right. 
\end{equation} 
The value for $A=2$ represents the \cite{Shu77} inside-out collapse solution, whereas the limit $A\gg 2$ is derived assuming pressureless collapse of each mass shell.   
It is possible to predict $\dot{M}$ analytically \citep{Shu77}, but in practice we initialize our simulations with a range of values $A>2$ and measure $\dot{M}$ just outside the disk.    
Because our equation of state is isothermal up to densities well above the typical disk density ($\csdisk= c_{s,{\rm core}}$),   $\dot{M} G/c_{s,{\rm core}}^3$ is equivalent to our parameter $\xi$. 

In order to set the value of our rotational parameter $\Gamma$ and hold it fixed, we initialize our cores with a constant, subsonic rotational velocity: 
\begin{equation}\label{omega}
\Omega = {2A c_s\over \pomega} \left(\frac{\Gamma}{\xi}\right)^{1/3} ,
\end{equation}
where $\pomega$ is the cylindrical radius.   We arbitrarily choose a constant velocity rather than rigid rotation on spheres in order to concentrate accretion near the outer disk radii.  Our definition of $\Gamma$ in terms of the mean value of $j_{\rm in}$ rather than its maximum value is intended to reduce the sensitivity of our results to the choice of rotational profile.  

 Given these initial conditions, our parameters $\xi$ and $\Gamma$ remain constant throughout the simulation, while the collapsed mass and disk radius (as determined by the Keplerian circularization radius of the infalling material) increase linearly with time. We define a resolution parameter,
 \begin{equation}
   \lambda = {R_{\rm k,in}\over dx_{\rm{min}}},
\end{equation}
to quantify the influence of numerics on our results.  Because we hold the minimum grid spacing $dx_{\rm min}$ constant, $\lambda$ increases $\propto t$ as the simulation progresses.

 By artificially controlling the infall parameters of our disks, and then watching them evolve in resolution, we gain insight into the physical behavior of accretion with certain values of $\xi$ and $\Gamma$, as captured in a numerical simulation with a given dynamical range ($\lambda$).  Our initial conditions are necessarily ideal, allowing us to perform controlled experiments: of course realistic star-forming cores will undoubtedly be somewhat turbulent with time variable infall rates. 
 
\subsection{Domain and Resolution} \label{domain}
Due to the dimensionless nature of these experiments, we do not use physical units to analyze our runs. The base computational  grid is $128^3$ cells, and for standard runs we use nine levels of refinement, with a factor of two increase in resolution per level: this gives an effective resolution of  $65,536^3$.  More relevant to our results, however, is the resolution with which our disks are resolved: $\lambda \lesssim 10^2$.   To compare this to relevant scales in star formation, this is equivalent to sub-AU resolution in disks of $\sim 50-100$ AU. 

The initial core has a diameter equal to one half of the full grid on
the base level. The gravity solver obeys periodic boundary conditions on the
largest scale; as the disk is 2.5 to 3 orders of magnitude smaller
than the grid boundaries, disk dynamics are unaffected by this choice.
 The initial radius of the current infall is $(\pi \Gamma)^{-2/3}
R_{\rm k,in}$ (from equations (\ref{gamma}), (\ref{iso_shu}), and (\ref{MdotCore})); although this is
much larger than the disk itself, it is still $\sim 15-40$ times
smaller than the initial core and $\sim 30-80$ times smaller than the
base grid.   Tidal distortions of the infall are therefore very small,
although they may be the dominant seeds for the GI.  We return to this
issue in \S \ref{reso}, where we compare two runs in which
only the tidal effects should be different.

In addition to the density criterion for grid refinement described in \S \ref{code}, we also refine spatially to ensure that the entire disk is resolved at the highest grid level. We use $\xi$ and $\Gamma$ to predict the outer disk radius (see \S \ref{scalings}), and refine to $150\%$ of this value in the plane of the disk. In the vertical direction, we refine to $40\%$ of the disk radius: this value is larger than the expected scaleheight for any of our disks by at least $\sim 15\%$. We find that we accurately capture the vertical and radial extent of the disk with this prescription, and the density criterion ensures that any matter at disk-densities extending beyond these radii will be automatically refined.

\subsection{Dynamical Self-Similarity}\label{selfsim_idea}

Because our goal is to conduct a parameter study isolating the effects of our parameters $\xi$ and $\Gamma$, we hold each fixed during a single run.   At a given resolution $\lambda$, we expect the simulation to produce consistent results regarding the behavior of the accretion disk, the role of the GI, and fragmentation into binary or multiple stars.   Since $\lambda$ increases linearly in time, each simulation serves as a resolution study in which numerical effects diminish in importance as the run progresses.   Because the GI is an intrinsically unsteady phenomenon, a disk should fluctuate around its mean values even when all three of $\Gamma$, $\xi$, and $\lambda$ are fixed.  Because of this, and because $\lambda$ changes over the run, we expect our runs to be self-similar, but only in a limited, statistical sense. 

To illustrate how we expect resolution to affect our results,  consider another unsteady 
problem: the turbulent wake of a solid body in air.  Our parameters $\xi$
and $\Gamma$ are analogous to the macroscopic parameters of the problem,
such as the body's aspect ratio and the Mach number of its motion,
whereas $\lambda$ is analogous to the Reynolds number of the
flow.  With all the parameters fixed, the flow field fluctuates around
a well-defined average state.   The average flow properties do depend
on Reynolds number, but  increasingly slowly in the high-Reynolds
limit.   In our simulations the resolution parameter is never fixed,
but increases linearly in time; therefore we expect the disk to settle
into an approximate steady state in which each ten orbits
resemble the previous ten.  Unlike a Sedov blast wave, we should \emph{not} expect our disks to be exactly invariant under scaling: this would not be consistent with the turbulent saturation of non-linear instabilities. 

Moreover, whereas many physical systems are captured perfectly in the limit of infinite resolution ($\lambda\rightarrow\infty$), this is not true of isothermal, gravitational gas dynamics, in which the minimum mass and spacing of fragments both scale as $\lambda^{-1}$ \citep{1992ApJ...388..392I}.   For this reason we quote the resolution $\lambda$ whenever reporting on the state of the disk-star system. 

We note that there exists a minimum scale in real accretion disks as well, namely the opacity-limited minimum fragment mass \citep{1976MNRAS.176..483R}.   The finite dynamical range of our numerical simulations is therefore analogous to a phenomenon of Nature, albeit for entirely different reasons. 

 \begin{figure*}
    \centering
 	\includegraphics[scale=0.8]{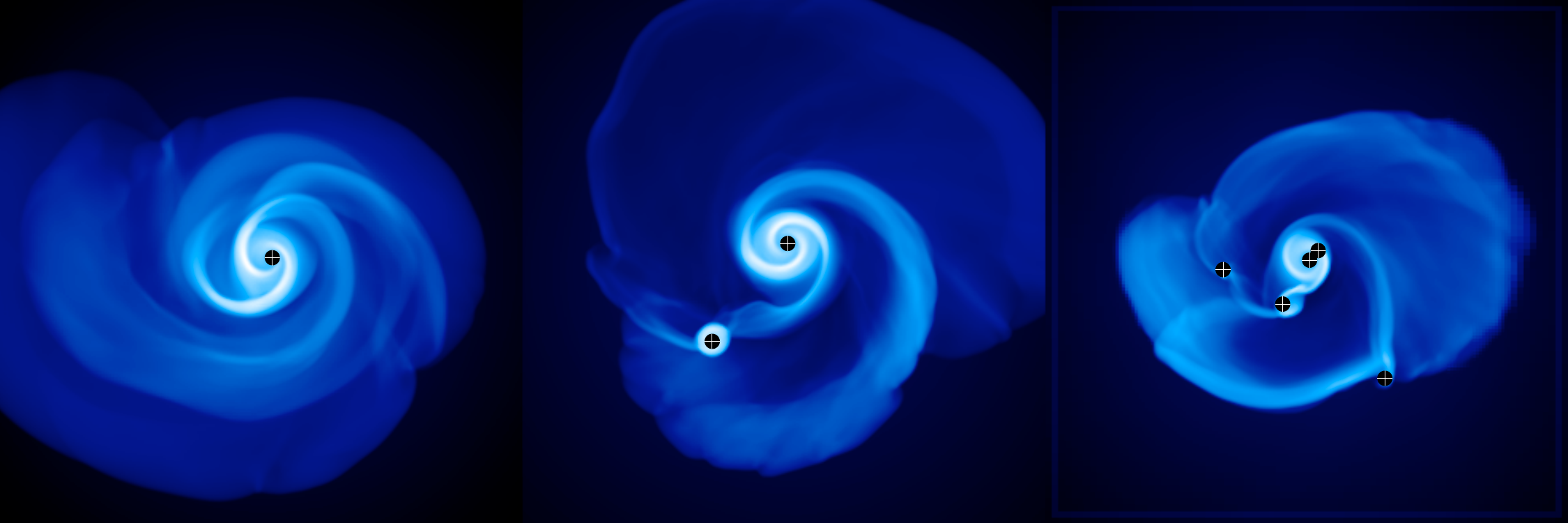}
  	\includegraphics[scale=0.8]{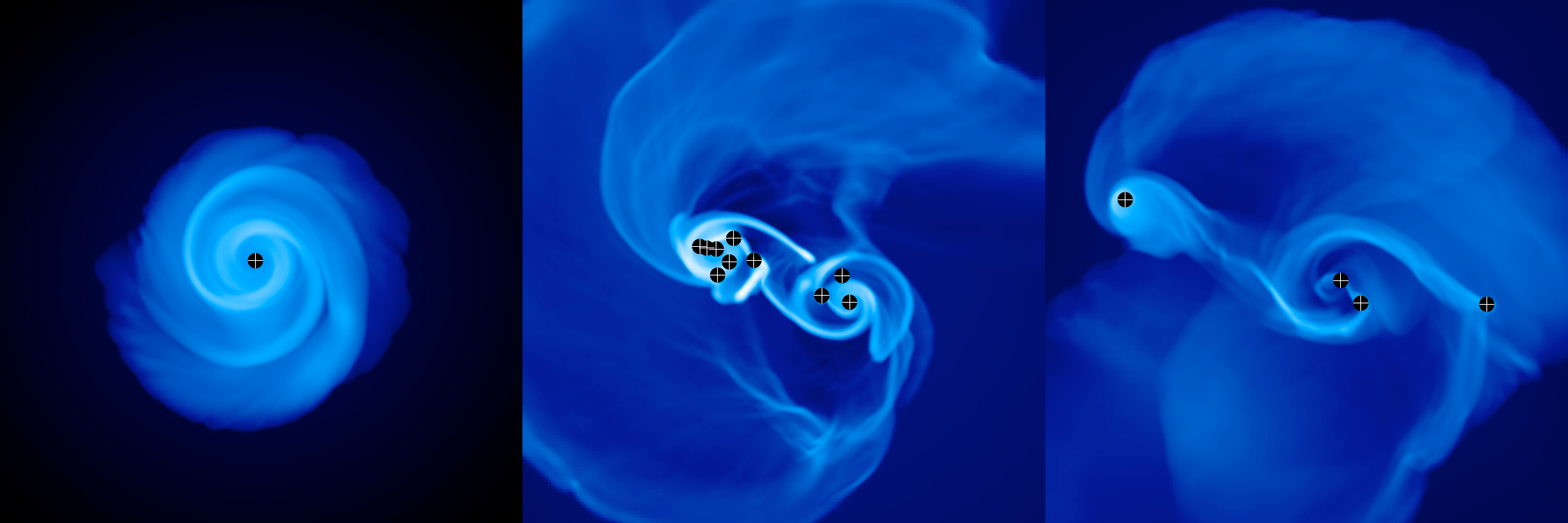}

    \caption{Two examples of single, binary, and multiple systems. The resolution across each panel is 328x328 grid cells. The single runs are $\xi=2.9, \Gamma=0.018$ (top), $\xi = 1.6, \Gamma = 0.009$ (bottom). The binaries are $\xi = 4.2, \Gamma = 0.014$ (top), $\xi = 23.4, \Gamma = 0.008$, (bottom). The multiples are $\xi = 3.0 , \Gamma = 0.016$ (top), $\xi = 2.4, \Gamma = 0.01$ (bottom). Black circles with plus signs indicate the locations of sink particles. These correspond to runs 5, 1, 9, 16, 7, and 4 respectively.}
    \label{prettypics}
 \end{figure*}

\section{Disk properties in terms of the accretion parameters} \label{scalings}
To assess the physical importance of $\xi$ and $\Gamma$, it is useful to consider the case of a single star and its accretion disk.  Because many $\xi$, $\Gamma$ pairs lead to fragmentation, this assumption is only self-consistent within a subregion of our parameter space; nevertheless it helps to guide our interpretation of the numerical results. In order to associate results from our parameters with those of previous studies, we also derive expressions for disk averaged quantities such as $Q$ and the disk-to-system mass ratio, $\mu$ as a function of $\xi$ and $\Gamma$. 

The combination 
\begin{equation} \label{HonRfromGamma/Xi} 
\left(\Gamma\over\xi\right)^{1/3} = {\left<j\right>_{\rm in} c_{s,d}\over G M_{*d} } = {c_{s,d}\over v_{\rm k,in}}
\end{equation} 
is particularly useful, since it provides an estimate for the disk's aspect ratio (the scale height compared to the circularization radius).  Being independent of $\Mdot$, it is more a property of the disk than of the accretion flow.  

The other important dimensionless quantity whose mean value depends primarily on $\xi$ and $\Gamma$ (and slowly on resolution) is the disk-to-system mass ratio 
\begin{equation}\label{mu}
\mu = {M_d\over M_{*d}}.
\end{equation}

When the disk is the sole repository of angular momentum,  the specific angular momentum stored in the disk is related to the infalling angular momentum via:
\begin{equation}\label{j_d_vs_j_in}
j_d = \left(J_{\rm in} \over \left<j\right>_{\rm in} M_{*d} \right) 
{\left<j\right>_{\rm in}\over \mu} 
\end{equation}
where $J_{\rm in}$ is the total angular momentum accreted, so that $J_{\rm in}/(\left<j\right>_{\rm  in} M_{*d}) = 1/(l_j+1)$ in an accretion scenario where $\left<j\right>_{\rm in}\propto M_{*d}^{l_j}$.   In our simulations $l_j=1$, so $j_d=\left<j\right>_{\rm  in}/(2\mu)$.  Given the relation between $j_d$ and $\left<j\right>_{\rm in}$, we can define
\begin{equation}\label{R_d_Omega_d}
\left.\begin{array}{c}R_d \\ \Omega_d   \end{array}\right.
\left.\begin{array}{c}= \\= \end{array}\right.
\left.\begin{array}{c} \left[(l_j+1)\mu\right]^{-2} R_{\rm k,in}  \\ \left[(l_j+1)\mu\right]^{3}\Omega_{\rm k, in}
 \end{array}\right.
\end{equation} 
which relate the disk's characteristic quantities ({\em not} the location of its outer edge) to conditions at the current circularization radius $R_{\rm k,in} = \left<j\right>_{\rm in}^2/(G M_{*d})$.  Such ``characteristic" quantities are valuable for describing properties of the disk as a whole, rather than at single location, with an effective mass weighting.  If we further suppose that the disk's column density varies with radius as $\Sigma(r) \propto r^{-k_\Sigma}$ (we expect $k_\Sigma\simeq 3/2$ for a constant Q, isothermal disk), we may define its characteristic column density $\Sigma_d = (1-k_\Sigma/2) M_d/(\pi R_d^2)$: 
\begin{equation}\label{Sigma_d}
\Sigma_d \simeq  f_\Sigma
{G^2 M_{*d}^3\over \left<j\right>_{\rm in}^4 }   
\mu^5
\end{equation}
where $f_\Sigma =   {(1-k_\Sigma/2) (1+l_j)^4 /\pi }$. Using equations (\ref{HonRfromGamma/Xi}) and (\ref{R_d_Omega_d})-(\ref{Sigma_d}), we can rewrite the Toomre stability parameter $Q$ (ignoring the difference between $\Omega$ and the epicyclic frequency for simplicity):
\begin{eqnarray}\label{Q_d}
Q & = & \frac{ c_s \kappa}{\pi G \Sigma} \rightarrow \frac{c_s \Omega_d}{\pi G \Sigma_d} \\
Q_d &\simeq& \frac{f_Q^{-1}}{\mu^{2}}  \frac{c_{s,d} \left<j\right>_{\rm in} }{G M_{*d} }
 \nonumber \\ 
 & &= \left(\frac{\Gamma}{\xi}\right)^{1/3} \frac{f_Q^{-1}}{ \mu^{2}}.
\end{eqnarray} 
 where $f_Q = (1-k_\Sigma/2)(1 + l_j) $.
 To the extent that we expect $Q_d\sim1$ in any disk with a strong GI, this suggests $\mu\sim (\Gamma/\xi)^{1/6} (1-k_\Sigma/2)^{-1/2}(1+l_j)^{-1/2}$; and because we expect that $\mu$ has an upper limit of around 0.5 \citep[see \S \ref{results} and discussion in KMK08 and][]{Sling1990}, we see there is an upper limit to $\xi/\Gamma$ above which the system is likely to become binary or multiple.  This is not surprising, as $\mu$ is proportional to scale height when $Q$ is constant; equation (\ref{Q_d}) simply accounts self-consistently for the fact that $\mu$ also affects $R_d$.   

To go any further with analytical arguments, we must introduce the \cite{SS1973} $\alpha$ viscosity parameterization, in which steady accretion occurs at a rate
\begin{equation}\label{Mdot_SS73} 
\Mdot_d(r) =  {3\alpha(r)\over Q(r)} {c_s(r)^3\over G}
\end{equation}  
 Using the definition of $\xi$
 \begin{equation} \label{xi_constrains_alpha/Q} 
\xi  \sim  {3\alpha(r) \over Q(r)}  {c_s(r)^3 \over  \csdisk^3} 
\end{equation}
 Insofar as $Q\sim1$ when the GI is active, the effective value of  $\alpha$ induced by a strong GI is directly proportional to $\xi$.   
  
The magnitude of $\Gamma$ has important implications for disk evolution.  As discussed previously by KMK08, $\Gamma$ (called $\Re_{\rm in}$ there) affects $\mu$ through the relation  
\begin{eqnarray}\label{muEvolution}
{\dot{\mu}\over \mu \Omega_{\rm k,in}}& =& \Gamma\left(\frac1\mu -1\right) - {\dot{M}_*\over M_d\Omega_{\rm k, in}}. \\  
& &\simeq  \Gamma\left(\frac1\mu -1\right) - 3(1-\frac{k_\Sigma}{2})(1+l_j)\alpha \mu\left(\Gamma\over\xi\right)^{2/3}, \nonumber 
\end{eqnarray} 
where the second line uses disk-averaged quantities to construct a mean accretion rate from equation (\ref{Mdot_SS73}).  In our simulations $\dot{\mu}\simeq 0$ so we expect $\mu$ to saturate at the value for which the two terms on the right of equation (\ref{muEvolution}) are equal, 
\begin{equation}\label{muSolution} 
\mu \rightarrow (B^2 + 2B)^{1/2} - B,  ~~{\rm where}~~ B = {\Gamma^{1/3} \xi^{2/3}\over 3(2-k_\Sigma) (1+l_j) \alpha}. 
\end{equation}  
The disk mass fraction $\mu$ increases with $B$, so both $\Gamma$ and $\xi$ have a positive effect on $\mu$, whereas $\alpha$ tends to suppress the disk mass.  Note that, when $B$ is small and $\mu\simeq \sqrt{2B}$, equation (\ref{Q_d}) implies $Q_d\simeq 3\alpha/\xi$ in accordance with equation (\ref{Mdot_SS73}).   Because the effective value of $\alpha$ induced by the GI is a function of disk parameters, we cannot say more without invoking a model for $\alpha(\Gamma, \xi)$ or $\alpha(Q,\mu)$ as in KMK08. 

The scalings of disk properties with the dimensionless parameters
of the problem are in accord with intuitive expectations. An
increase in $\xi$ corresponds to an increase in accretion rate at
fixed disk sound speed, and as a result the equilibrium disk mass
rises. An increase in $\Gamma$ corresponds to an increase in the mean
angular momentum of the infall at fixed sound speed, leading to larger
disks that must transport more angular momentum, and thus again become
more massive. An increase in $\alpha$ corresponds to an increase in
the rate at which the disk can transport angular momentum and mass at
a fixed rate of mass and angular momentum inflow, allowing the disk to
drain and reducing its relative mass.
   We use the above relations to guide our interpretation of our simulation results, specifically the dependence of disk parameters like $\mu$, $Q_d$, $\alpha$, and the fragmentation boundary, on $\xi$ and $\Gamma$. 

\section{Results}\label{results}

Each of our simulations produces either a disk surrounding a single star, or binary or multiple star system formed via disk fragmentation; Figure \ref{prettypics} depicts examples of each outcome.  We use these three possible morphologies to organize our description of the simulations.  We explore the properties of each type of disk below as well as examine the conditions at the time of fragmentation. 

\begin{figure}
    \centering
 	\includegraphics[scale=0.5]{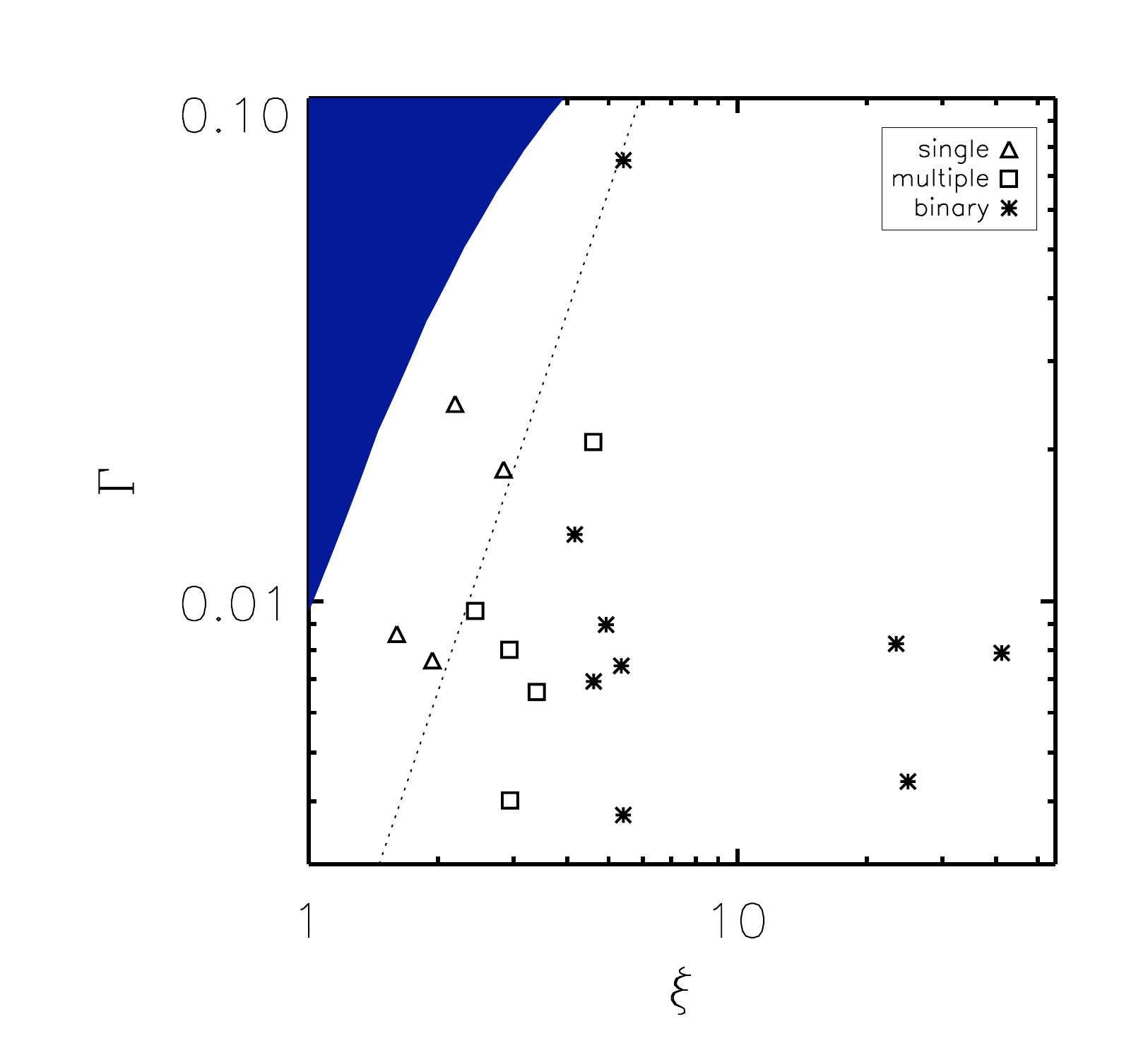}
 
    \caption{ Distribution of runs in $\xi- \Gamma$ parameter space. The single stars are confined to the low $\xi$ region of parameters space, although increasing $\Gamma$ has a small stabilizing effect near the transition around $\xi = 2$ due to the increasing ability of the disk to store mass at higher values of $\Gamma$. The dotted line shows the division between single and fragmenting disks: $\Gamma = \xi^{2.5}/850$. As $\xi$ increases disks fragment to form multiple systems. At even higher values of $\xi$ disks fragment to make binaries. We discuss the distinction between different types of multiples in \S \ref{binstuff}. The shaded region of parameter space shows where isothermal cores no longer collapse due to the extra support from rotation. }
    \label{xi_gamma}
 \end{figure}

The division between single and fragmenting disks in $\xi$ and $\Gamma$ is relatively clear from our simulation results, as shown in 
Figure \ref{xi_gamma}. Several trends are easily identified. First, there is a critical $\xi$ beyond which disks fragment independent of the value of $\Gamma$. Below this critical $\xi$ value, there is a weak stabilizing effect of increasing $\Gamma$. As $\xi$ increases, disks transition from singles in to multiples, and finally into binaries. We discuss the distinction between binaries and multiples in \S \ref{binstuff}.  
  
  In table \ref{restab} we list properties of the final state for all of our runs, their final multiplicity (S, B, or M for single, binary, or multiple, respectively), and the disk-to-star mass ratio $\mu_f$ measured at the time at which we stop each experiment, as well as the maximum resolution $\lambda_n$.   Note that the disk extends somewhat beyond $R_{\rm k,in}$: therefore the disk as a whole is somewhat better resolved than the value of $\lambda_n$ would suggest.  For the disks which fragment, we also list the value of $\mu_f, \lambda_f$ and $Q$ just before fragmentation occurs.

\begin{table}[htdp] 
\begin{center}
\begin{tabular}{ c l c c c c c c c c |} 
\hline
\hline
 $\#$ & $\xi$ & $ {10^2} \Gamma$ & $N_*$ & $\mu_f$ &$\lambda_f$&$Q_{2D}$ & $\mu $ & $\lambda_n$ \\

\hline
1&1.6 & 0.9 & S & ... &...&...&0.49&99\\
2& 1.9 & 0.8 & S & ...&...&...&0.40&88\\
3&2.2 & 2.5 & S & ...& ... &...&0.56&82\\ 
4&2.4 & 1.0 &  M & 0.43 &77&0.69&0.16&98\\
5&2.9 & 1.8 & S & ... &...&...&0.53&86\\
6& 2.9 & 0.8 & M & 0.40&51&0.72&0.14&78 \\
7&3.0 & 0.4 & M & 0.33 &50&0.48&0.11&77\\
8&3.4 & 0.7 & M & 0.40 &66&0.37&0.16&70\\
9&4.2 & 1.4 & B & 0.51 &56&0.19&0.33&72\\
10&4.6 & 2.1 & M & 0.54 &71&0.42&0.23&123\\
11&4.6& 0.7 & B & 0.35 &28&0.52&0.12&52\\
12&4.9 & 0.9 & B & 0.37&26&0.74&0.19&59 \\
13&5.4 & 0.4 & B & 0.38&38&0.33&0.19&64\\ 
14& 5.4 & 0.7 & B & 0.31 &49&0.85&0.21&62\\
15&5.4 & 7.5 & B & 0.72 &99&0.20&0.59&129\\
16*&23.4 & 0.8 & B & 0.25 &5&0.83&0.10&84\\
17*&24.9 & 0.4 & B & 0.15 &3&0.59&0.11&61\\
18*&41.2 & 0.8 & B & 0.13 &5&1.33&0.10&58\\

\hline
\end{tabular} 
\caption{Each run is labelled by $\xi, \Gamma$, multiplicity outcome, the final value of the disk-to-star mass ratio,$\mu$ and the final resolution, $\lambda_n$. Values of $\Gamma$ are quoted in units of $10^{-2}$. For fragmenting runs the disk resolution $\lambda_f$, $Q_{2D}$ (equation \ref{Q_estimators}) and $\mu_f$ at the time of fragmentation are listed as well. S runs are single objects with no physical fragmentation. B's are binaries which form two distinct objects each with a disk, and M are those with three or more stars which survive for many orbits. *  indicates runs which are not sufficiently well resolved at the time of fragmentation to make meaningful measures of $\mu_f$, and $Q$. }

\label{restab}
\end{center}

\end{table}%
 
In table \ref{singletab} we describe those disks which do not fragment: we list the analytic estimate for the characteristic value Toomre's $Q$,  $Q_d$, the measured minimum of $Q_{2D}$ (equation \ref{Q_estimators}), the radial power law $k_\Sigma$ which characterizes $\Sigma(r)$ for a range of radii extending from the accretion zone of the inner sink particle to the circularization radius $R_{\rm k,in}$, the final disk resolution, $\lambda_n$, and the characteristic disk radius, $R_d$ (equation (\ref{R_d_Omega_d}).

 \begin{table}[htdp] 
\begin{center}
\begin{tabular}{c l c c c c c c c c |} 
\hline
\hline
 $\#$ & $\xi$ & ${10^2}\Gamma$ & $\mu $ &$Q_d$&$Q_{2D}$&$k_\Sigma$ &$\lambda_n$& $R_d$ \\

\hline
1&1.6 & 0.9 &0.49&1.6&0.96&1.5&99 &103\\
2& 1.9 & 0.8 & 0.40&1.5&1.10&1.3&88&138\\
3&2.2 & 2.5 &  0.56 & 3.7&0.83&1.8&82&65\\ 
5&2.9 & 1.8 & 0.53&2.2&0.56&1.7&86&77\\
\hline
\end{tabular} 
\caption{Single runs (numbers as from table \ref{restab}). We list values for the characteristic predicted value of Toomre's $Q$,  $Q_d$ (equation \ref{Q_d}),  as well as the measured disk minimum, $Q_{2D}$ equation (\ref{Q_estimators}). We also list the slope of the surface density profile, $k_\Sigma$ averaged over several disk orbits, the final resolutions, and $R_d$ at the end of the run (equation \ref{R_d_Omega_d})}

\label{singletab}
\end{center}
\end{table}
 
\subsection{The Fragmentation Boundary and $Q$}\label{Qstuff}
\begin{figure}
   \centering
   \includegraphics[scale=.7]{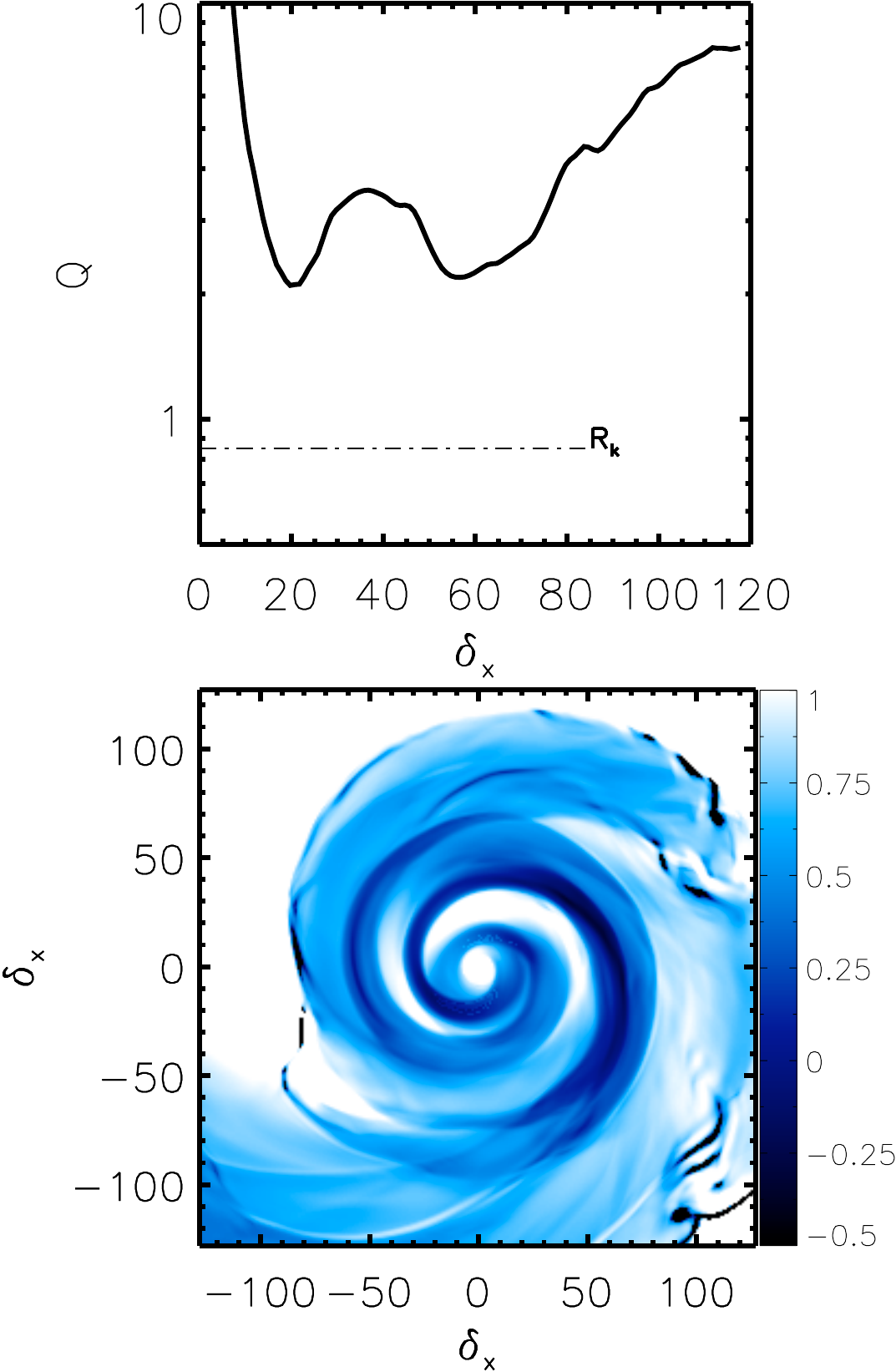} 
   \caption{Top: $Q_{\rm av}$ in a disk with $\xi = 2.9, \Gamma = 0.018$. The current disk radius, $R_{\rm k,in}$ is shown as well. Bottom: Log($Q_{2D}$) (equation \ref{Q_estimators}) in the same disk. While the azimuthally averaged quantity changes only moderately over the extent of the disk, the full two-dimensional quantity varies widely at a given radius. $Q$ is calculated using $\kappa$ derived from the gravitational potential, which generates the artifacts observed at the edges of the disk. Here and in all figures, we use $\delta_x$ to signify the resolution.}
   \label{Qaz}
\end{figure}

It is difficult to measure a single value of $Q$ to characterize a disk strongly perturbed by GI, so we consider two estimates: a two dimensional measurement $Q_{\rm 2D}$, and a one-dimensional measure $Q_{\rm av}(r)$ based on azimuthally-averaged quantities.
\begin{eqnarray}\label{Q_estimators} 
Q_{\rm 2D} (r,\phi)&= &\frac{ c_s \kappa}{\pi G \Sigma},\\
Q _{\rm av}(r) &= &\frac{\bar{c}_s(r)\bar{\kappa} (r) }{\pi G \bar{\Sigma}(r)} 
\end{eqnarray}
(bars represent azimuthal averages). 
As Figure \ref{Qaz} shows, the two-dimensional estimate shows a great deal of structure which is not captured by the azimuthal average, let alone by $Q_d$.   Moreover, while the minimum of the averaged quantity is close to two, the two dimensional quantity drops to $Q \sim 0.3$.  We find that the best predictor of fragmentation is the minimum of a smoothed version of the two-dimensional quantity (smoothed over a local Jeans length to exclude meaningless fluctuations), although $Q_d$ shows a similar trend. We use this quantity in table \ref{restab}, and compare it to the analytic estimate $Q_d$ in table \ref{singletab} for non-fragmenting disks.

 We emphasize that the critical values of $Q$ at which fragmentation sets in depend on the exact method used for calculation (e.g. $Q_{\rm az}$ or $Q_{2D}$).  Moreover, we do not expect to reproduce fragmentation at the canonical order unity boundary. This only marks the critical case for the m=0 unstable mode in razor-thin disks \citep{Toom1964}. As discussed by numerous authors, the fragmentation criterion is somewhat different for thick disks \citep{GLB1965,Laughlin:1997fk,1998ApJ...504..945L}, and the growth of higher order azimuthal modes \citep{ ARS89,Sling1990,Laughlin:1996uq}. 

\begin{figure}
    \centering
 	\includegraphics[scale=0.5]{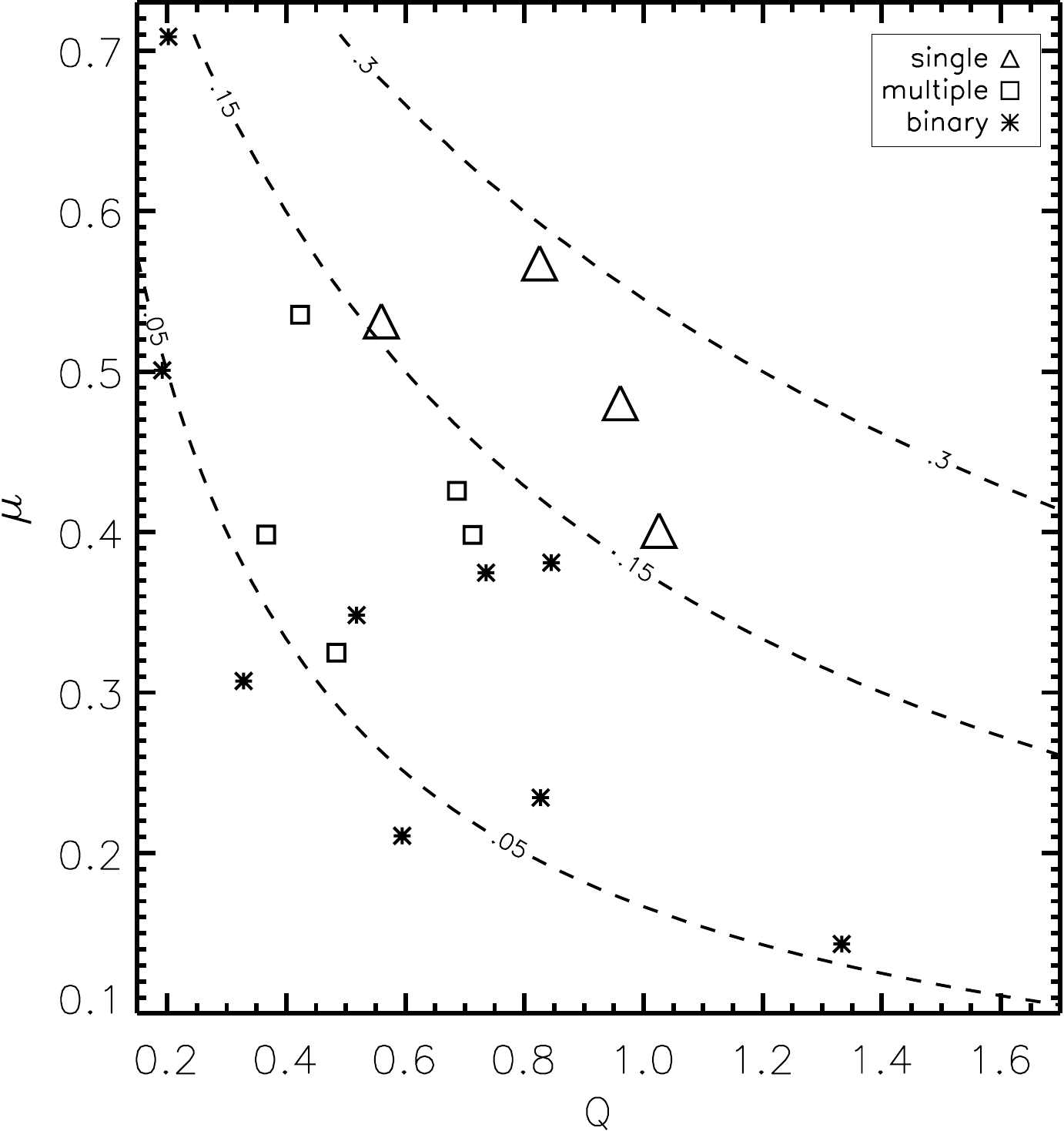}
 
    \caption{ Steady-state and pre-fragmentation values of $Q$ and $\mu$ for single stars and fragmenting disks respectively. We use the minimum of $Q_{2D}$ as described in \S \ref{Qstuff}. Symbols indicate the morphological outcome. Note that the non-fragmenting disks (large triangles) have the highest value of $\mu$ for a given $Q$. Contours show the predicted scaleheight as a function of $Q$ and $\mu$. It is clear that the single disks lie at systematically higher scale heights. We have assumed $k_\Sigma = 3/2$ in calculating scaleheight contours as a function of $Q$ and $\mu$.}
    \label{q_mu}
 \end{figure}

Another  consequence of trying to describe thick disks with multiple unstable modes is that the fragmentation boundary cannot be drawn in $Q$-space alone. We use $Q_{2D}$ and $\mu$ in Figure \ref{q_mu} to demarcate the fragmentation boundary. Labeled curves illustrate that the critical $Q$ for fragmentation depends on the disk scale height (equation \ref{HonRfromGamma/Xi}). At a given value of $Q$, a disk with a larger value of $\mu$ will have a larger aspect ratio, and will therefore be more stable.  Recall from equation (\ref{HonRfromGamma/Xi}), that the disk aspect ratio is proportional to $\left(\xi/\Gamma\right)^{1/3}$. 

 This trend is consistent with the results of  \cite{GLB1965}  for thick disks; because the column of material is spread out over a larger distance, $H$, its self-gravity is somewhat diluted.  The fact that two parameters are necessary to describe fragmentation is also apparent in Figure \ref{xi_gamma}, where the boundary between single and multiple systems is a diagonal line through the parameter space. 
 
\begin{figure*}
    \centering
\includegraphics[scale=0.8]{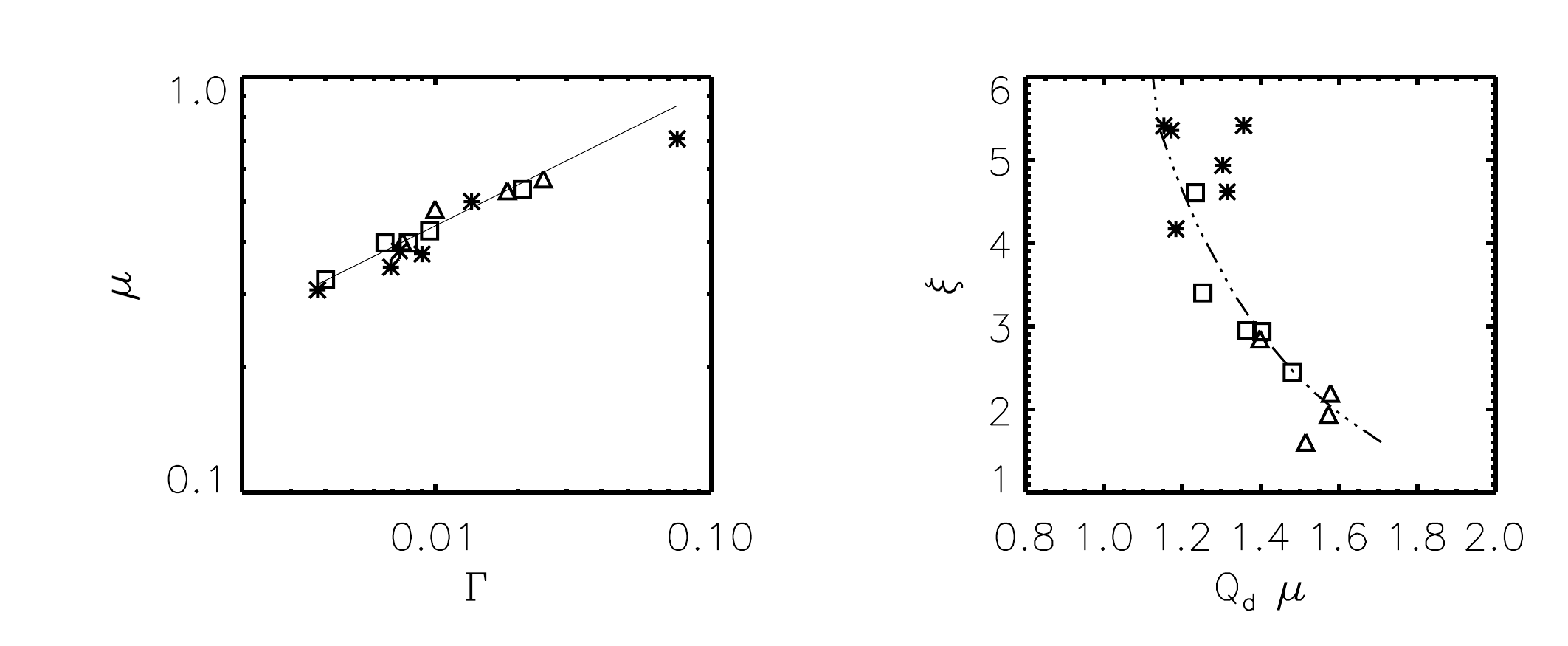}
    \caption{At right $\Gamma$ vs $\mu$ with the fit in equation (\ref{mugam}) overplotted. At left, $Q_d \mu$ vs $\xi$ with the scaling $ Q \propto \xi^{-1/3}$ overplotted. Runs, 16, 17, 18 are omitted as the low resolution at the time of fragmentation makes measurements of $\mu$ and therefore $Q_d$ unreliable. }
    \label{mu_gam_comp}
 \end{figure*}

 Although two criteria are necessary to prescribe the fragmentation boundary, we observe a direct correspondence between $\mu$ and $\Gamma$, and $\xi$ and Toomre's $Q$. Figure \ref{mu_gam_comp} shows that $\mu \approx 2\Gamma^{1/3}$ for both single star disks, and just prior to the onset of fragmentation in disks that form binaries and multiples. We find a similar correspondence between $\xi$ and the combination $Q_d \mu  $, which is a direct correlation between $\xi$ and $Q$ defined with respect to the disk circularization radius (using $R_d$ in the definition of $Q_d$ brings in an extra factor of $ \mu$.) 

\subsection{Properties of non-fragmenting disks}\label{mustuff}

\begin{figure}
   \centering
   \includegraphics[scale=0.55]{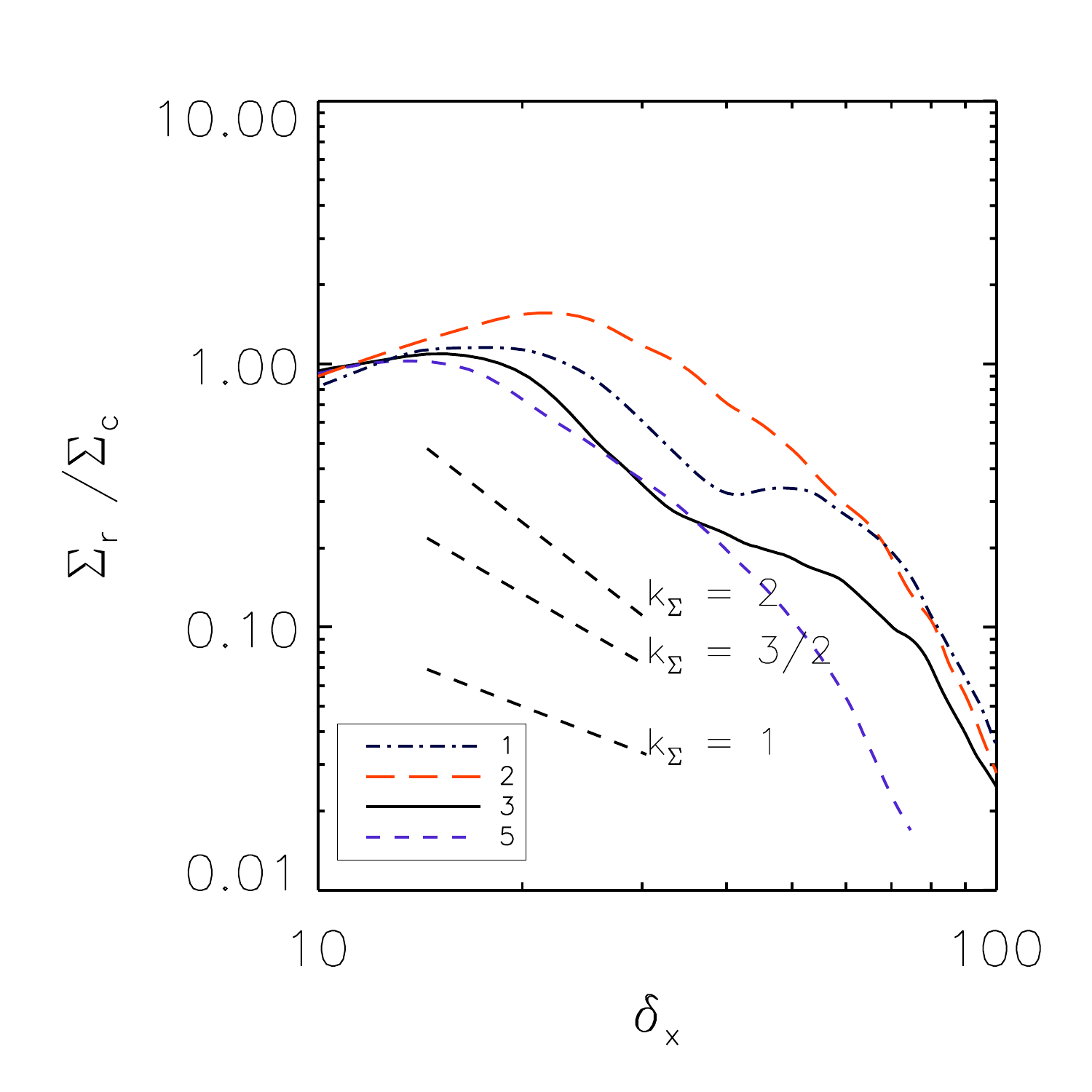} 
   \caption{Normalized density profiles for the single-star disks. Profiles are azimuthal averages of surface densities over the final $\sim 3$ disk orbital periods. We find that while the inner regions are reasonably approximated by power law slopes, the slope steepens towards the disk edge. For comparison, slopes of $k_\Sigma= 1,1.5,$ and $ 2$ are plotted as well. Runs are labelled according to their values in table \ref{restab}. }
   \label{sigmaprofs}
\end{figure}

Although we quote a single power law value for the surface density profiles of disks in table \ref{singletab}, the surface density structure is somewhat more complex. We find that the disks show some evidence of a broken power law structure: an inner region, characterized by $k_\Sigma$, where disk material is being accreted inwards, and an outer region characterized by a steep, variable power law due to the outward spread of low-density, high angular momentum material. We find disks characterized by slopes between  $k_\Sigma = 1 - 2$. Clustering around $k_\Sigma = 3/2$ is expected, as this is the steady-state slope for a constant $Q$, isothermal disk. Our measurements of $Q(r)$ (equation \ref{Q_estimators}) show fluctuating, but roughly constant value over the disk radius.  Note that the slope of the inner disk region tends to increase with $\Gamma$. Figure \ref{sigmaprofs} shows normalized radial profiles for  the non-fragmenting disks. Profiles are averaged over approximately three disk orbital periods. The flattening at small radii is due to the increasing numerical viscosity in this region (\S \ref{reso}). 

We find an upper mass limit of $\mu  \sim 0.55$, for single stars, which means that disks do not grow more massive than their central star. A maximum disk mass has been predicted by \cite{Sling1990} as a consequence of the SLING mechanism. Such an upper limit is expected as eccentric gravitational instabilities in massive disks shift the center of mass of the system away from the central object. Indeed, we observe this wobble in binary forming runs. The subsequent orbital motion of the primary object acts as an indirect potential exciting strong $m=1$ mode perturbations which can induce binary formation \citep{Sling1990}. We find that this maximum value is consistent with their prediction. 

Using the analytic expressions above, we can also derive an expression for an effective Shakura-Sunyaev $\alpha$. In this regime of parameter space,  $\xi$ and $\Gamma$ are always such that $B \ll1$ (assuming $\alpha$ does not stray far from unity). We therefore expect that $\mu \propto \Gamma^{1/6}\xi^{1/3}\alpha^{-1/2}$. Using this relation we can find a functional form of $\alpha(\xi,\Gamma)$. Our fit to the data shown in Figure \ref{mu_gam_comp}  implies 

\begin{equation}\label{mugam}
\mu \approx 2 \Gamma^{1/3},
\end{equation}
with some scatter for both single disks and fragmenting disks just prior to fragmentation. We can use this fit to infer a scaling relation for $\alpha$ using equation (\ref{muSolution}) in the limit $\mu \sim \sqrt{2B}$:
\begin{equation}\label{alphascale}
\alpha_d \approx \frac{1}{18(2-k_\Sigma)^2(1+l_j)^2} \frac{\xi^{2/3}}{\Gamma^{1/3}}.
\end{equation}

The scaling is consistent with our expectation that driving the disk with a higher $\xi$ causes it to process materially more rapidly, while increasing $\Gamma$ decreases the efficiency with which the disk accretes. Equation (\ref{alphascale}) predicts disk averaged values of $\alpha$ for single star disks between $\sim 0.3-0.8$.  These values are consistent with the observed accretion rates, and numerically calculated torques (\S \ref{torques}).

\subsection{ The formation of binaries and multiples}\label{binstuff}
As shown by Figure \ref{xi_gamma}, a large swath of our parameter space is characterized by binary and multiple formation. We find that the division between fragmenting and non-fragmenting disks can be characterized by a minimum value of $\Gamma$ at which disks of a given $\xi$ are stable.  In Figure \ref{xi_gamma} we have plotted this boundary as $\Gamma = \xi^{2.5}/850$. 

While we do not claim that our numerical experiments are a true representation of the binary formation process, we do expect to find binaries in much of the parameter space characteristic of star formation, as nearly half of all stars are in binaries \citep{1991A&A...248..485D,2007prpl.conf..379D}. Moreover, as the binary forming parameters are typical of higher mass star formation, where binaries and multiples are expected to comprise perhaps 75\% of systems, these findings are encouraging \citep{1998AJ....115..821M, 2009AJ....137.3358M}.  We discuss several general trends here, but defer a detailed analysis of binary evolution and application to observations to a later paper. 

Are these equal mass binaries? Low mass stellar companions? Or maybe even massive planets? In a self-similar picture it is difficult to tell. In an actively accreting multiple system, as long as the mass reservoir has angular momentum such that the circularization radius of the infalling material is comparable to the separation between objects, the smaller object, which is further from the center of mass, will accrete due to the torque imbalance \citep{Bate97,1994MNRAS.271..999B}. Similarly, in thick, gravitationally unstable disks, the isolation mass approaches the stellar mass:
\begin{equation}
 M_{\rm iso} = 4 \pi f_H r_H r_d \Sigma \approx 30 \frac{f_H}{3.5} \left(\frac{H}{R}\right)^{3/2}Q^{-3/2}M_*. 
 \end{equation}
 Here, $r_H = (M_s / 3 M_*)^{1/3}$ is the Hill radius, $M_s$, and $M_*$ are the masses of the secondary and primary, and the numerical factor $f_H$, represents how many Hill radii an object can feed from in the disk -- numerical simulations suggest $f_H \sim 3.5$ \citep{1987Icar...69..249L, 2002ApJ...572..566R}.  Therefore the evolution of these objects in our models is clear: they tend to equalize in mass. The binary separation will also grow if any of the infalling angular momentum is transferred to the orbits as opposed to the circumstellar disks. These trends are borne out in our experiments: binary mass ratios asymptote to values of $~0.8-0.9$ and separations to $\sim 60\%$ of $R_{\rm{k,in}}$.  In a realistic model for star formation, the parameters that characterize a single run in this paper will represent only one phase in the life of a newborn system. The trajectory through $\xi-\Gamma$ space which the systems take following binary formation will strongly influence the outcome in terms of separation and mass ratio. For example, should the disk stabilize and accretion trail off quickly following binary formation, it is quite likely that a large mass ratio would persist as the disk drains preferentially onto the primary object once the secondary reaches its isolation mass. By contrast, in systems which fragment before most of the final system mass has accreted, we expect more equal mass ratios. 

\subsubsection{ Hierarchical multiples and  resolution dependence}
Disks which are at the low $\xi$ end of the binary forming regime tend to form binaries at later times, and therefore at higher disk resolution. One consequence of this is the formation of hierarchical multiples. When disks become violently unstable, they fragment into multiple objects. Because of the numerical algorithm which forces sink particles within a gravitational softening length of each other to merge, at lower resolution many of these particles merge, leaving only two distinct objects behind. At higher resolution, while some of the particles ultimately merge, we find that three or four objects typically survive this process.  We cannot distinguish between merging and the formation of very tight binaries.  In addition to merging, small mass fragments are occasionally ejected from the system entirely. This appears be a stochastic process, though we have not done sufficient runs to confirm this conclusion. 

Disks which form binaries at early times and develop two distinct disks can also evolve into multiples when each disk becomes large enough and sufficiently unstable to fragmentation. In general, once a binary forms, the system becomes characterized by new values of $\xi$ and $\Gamma$ which are less than those in the original disk. As the distribution of mass and angular momentum evolves in the new system, the relative values of $\xi$ and $\Gamma$ evolve as well. However, once the mass ratios have reached equilibrium as is the case for run \#16 shown in the bottom center of Figure \ref{prettypics},  each disk sees $\xi$ of roughly half the original value, which for an initial $\xi \sim 24$ is still well into the fragmenting regime. As a result, the fact that the two disks ultimately fragment is expected. On the contrary, for the lowest $\xi$ binary runs, once one fragmentation event occurs, the new $\xi$ may be sufficiently low to suppress further fragmentation. The evolution of $\Gamma$ in the newly formed disks is more complicated, depending on how much angular momentum is absorbed into the orbit as compared to the circumstellar disks. We defer a discussion of this to a later paper.

 It is clear that there is a numerical dependence to this phenomenon which we discuss in \S \ref{reso}, but there is a correspondence with the physical behavior of disks as well. The radius and mass of a fragmenting disk are likely to influence the multiplicity outcome of a real system. Cores with high values of $\xi$ that form binaries early in our numerical experiments correspond to cores whose disks fragment into binaries at small physical size scales, where the disk may only be a few fragment Hill radii wide, and contain a relatively small number of Jeans masses. It is possible that at these size scales, numerous bound clumps in a disk might well merge leaving behind a lower multiplicity system than one in which the ratio of disk size to Hill radii or mass to Jeans mass is higher.

 \begin{figure}
     \centering
 \includegraphics[scale=0.60]{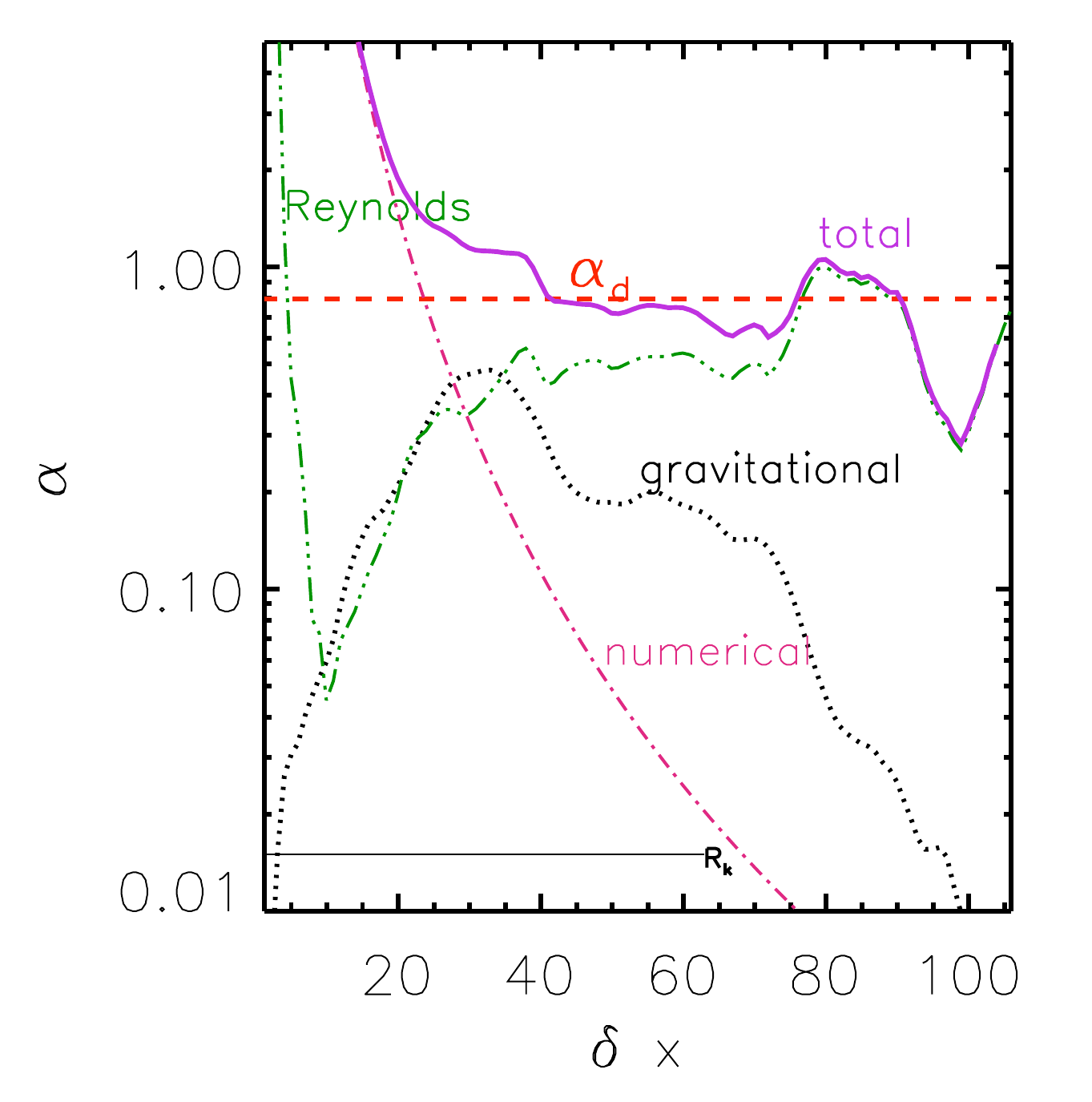}
  \caption{ Azimuthal averages of different components of torque expressed as an effective $\alpha$ (equation \ref{stressalpha}) for run \#8. The straight line, $\alpha_d$ (equation \ref{alphascale}) is plotted for comparison. The agreement between the analytic value of $\alpha_d$ and the combined contribution from the other components is best near the expected disk radius $R_{\rm k,in}$.}
    \label{azimalpha}
 \end{figure}

\subsection{ Gravitational Torques and Effective $\alpha$} \label{torques}
We verify that the accretion observed in our disks is generated by physical torques by computing the net torque in the disk. 
It is convenient to analyze the torques in terms of the stress tensor, $T_{R\phi}$, which is made up of two components: large scale gravitational torques and Reynolds stresses. Following \cite{LodRi05} we define:

\begin{equation} \label{stresstens}
T_{R\phi} = \int{\frac{g_R g_\phi}{4 \pi G} dz} + \Sigma \delta \vecv_R \delta \vecv_\phi,
\end{equation}

where $\delta \vecv = \vecv - \bar{v}$. In practice, we set $\delta \vecv_R = \vecv_R$, while $\delta \vecv_\phi$ is calculated with respect to the azimuthal average of the rotational velocity at each radius. In reality there is an extra viscous term attributable to numerical diffusion. We discuss the importance of this term in \S \ref{reso}.

The first term in equation (\ref{stresstens}) represents torques due to large scale density fluctuations in spiral arms, while the second is due to Reynolds stresses from deviations in the velocity field from a Keplerian (or at least radial) velocity profile.  To facilitate comparison with analytic models, the torques can be represented as an effective $\alpha$ where:
\begin{equation}\label{stressalpha}
T_{R\phi} = \left|{\frac{{\rm d ln}\Omega}{{\rm d ln} R}}\right| \alpha \Sigma c_s^2
\end{equation}

We can compare these torques to the characteristic disk $\alpha_d$ in equation (\ref{alphascale}). Although there is variability in the disk accretion with time, it is consistent with a constant rate over long timescales.

Figure \ref{azimalpha} compares  $\alpha_d$ to the azimuthal average of the physical torques for one of our runs. We also show the expected contribution from numerical diffusion (see \S \ref{reso}). The accretion expected from these three components is consistent with the time averaged total accretion rate onto the star. Due to the short term variability of the accretion rate, the two do not match up exactly.  It is interesting to note the radial dependence of the Reynolds stress term, which in the inner region decays rapidly, before rising again, due to the presence of spiral arms. In both the azimuthal average and the two dimensional distribution we see that at small radii numerical diffusion dominates, whereas at large radii deviations in the azimuthal velocity which generate Reynolds stresses are spatially correlated with the spiral arms. 

\subsection{ Vertical Structure} \label{vertical_struct}
\begin{figure}
    \centering
 	\includegraphics[scale=0.7]{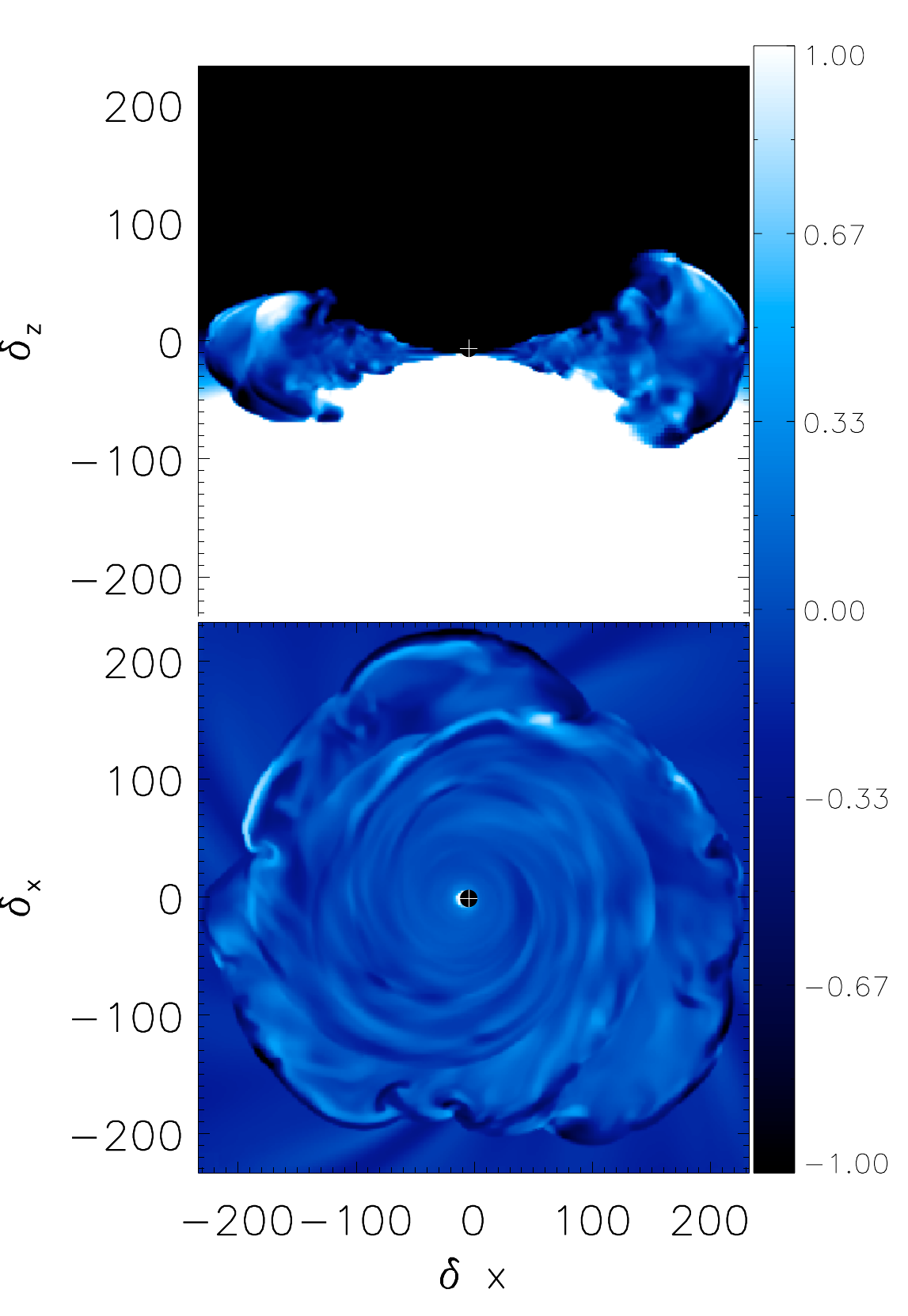}
 
    \caption{Cuts along the vertical axis and disk midplane of the vertical velocity, normalized to the disk sound speed. Clearly most of the vertical motions in the disk are transonic, although at the edges of the disk the velocities exceed $\mach \sim 1$. }
    \label{non-super}
 \end{figure}

When the disks reach sufficient resolution, we can resolve the vertical motions and structure of the disk. We defer a detailed analysis of the vertical structures to a later paper, but discuss several general trends here. Depending on the run parameters, the disk scale height is ultimately resolved by 10-25 grid cells. We observe only moderate transonic motions in the vertical direction of order $\mathcal{M} \sim 1-2$. Figure \ref{non-super} shows two slices of the z-component of the velocity field for a single system, one through the X-Z plane, and the other through the disk midplane. Although there is significant substructure, the motions are mostly transonic. 

\begin{figure}
    \centering
 	\includegraphics[scale=0.5]{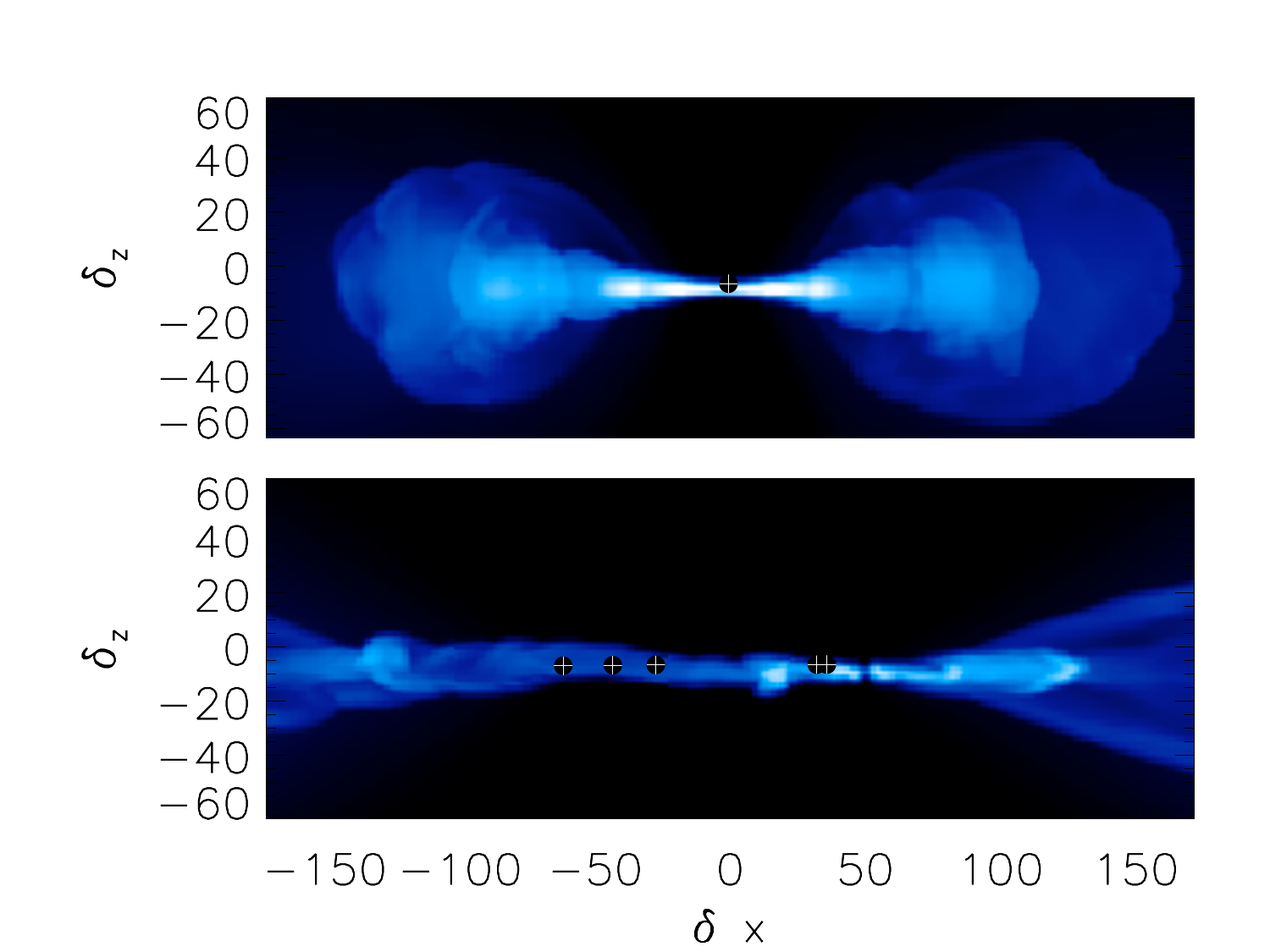}
 
    \caption{Density slices showing vertical structure in a single and binary disk. The top plot is a single star with $\xi = 1.6, \Gamma = 0.09$, while the bottom is a fragmenting binary system with $\xi = 24.3, \Gamma = 0.008$. The extended material in the binary system is generated by a combination of large scale circumbinary torques and the infalling material.  Colorscale is logarithmic.  The box sizes are scaled to $1.5 R_{\rm k,in}$ in the plane of the disk.}
    
    \label{scaleheight}
 \end{figure}

We also observe a dichotomy in the vertical structure between single and binary disks. Although the values of $\xi$ and $\Gamma$ should dictate the scaleheight (see equation (\ref{scaleheight})), and therefore higher $\xi$ disks which become binaries should have smaller scaleheights to begin with, we observe a transition in scaleheight when a disk fragments and becomes a binary.  Large plumes seen in single disks, like those shown in Figure \ref{scaleheight} contain relatively low density, high angular momentum material being flung off of the disk. The relatively sharp outer edges are created by the accretion shock of infalling material onto these plumes. We observe small scale circulation patterns which support these long lived structures.  Disks surrounding binaries, by comparison remain relatively thin; in particular while the circumprimary disks are slightly puffier than expected from pure thermal support, the circumbinary disk (when present) is sufficiently thin that we do not consider it well resolved. This implies that the effective $\Gamma$ values that binary disks see declines more than $\xi$ according to equation (\ref{HonRfromGamma/Xi}). This is consistent with the statement that some of the infalling angular momentum is transferred into the orbit instead of on to the disks themselves.

\section{ Caveats and Numerical Effects}\label{caveats}

\subsection{Isothermal equation of state} \label{thermo}

Many simulations have shown the dramatic effects that thermodynamics have on disk behavior \citep{2000ApJ...528..325B, Gam2001, Rice05, LodRi05,  2006ApJ...651..517B, Krumholz2007a, {2009arXiv0904.2004O}}. Since we are concerned with fragmentation, we must be aware of the potential dependencies of the fragmentation boundary on cooling physics. Starting with \cite{Gam2001}, there has been much discussion of the ``cooling time constraint" that states that a disk with $Q \sim 1$ will only fragment if the cooling time is short. While this is a valuable analysis tool for predicting the evolution of a system from a snapshot and for quantifying the feedback from gravito-turbulence, for most of the protostellar disks that we are modeling, the cooling time at the location of fragmentation is short. In the outer radii of protostellar disks in general, irradiation is the dominant source of heating \citep{1997ApJ...474..397D,ML2005,Krumholz2007a,KMK08}. In fact, even a low temperature radiation bath can contribute significantly to the heat budget of disks \citep{2008ApJ...673.1138C}. Such passively heated (through irradiation) disks behave more like isothermal disks than barotropic disks, because the energy generation due to viscous dissipation is small compared to the energy density due to radiation. Consequently, feedback from accretion in the midplane does not alter the disk temperature significantly. For realistic opacity laws, disks which are dominated by irradiation cool on timescales much faster than the orbital period at large radii. Numerical simulations such as \cite{Krumholz2007a} find that strongly irradiated disks have a nearly isothermal equation of state. In fact, the morphological outcome is similar to those of \cite{Krumholz2007a} with comparable values of $\xi$.

Another possible concern is the lack of a radial temperature gradient, independent of the equation of state. Both passively  and actively (through viscous dissipation) heated disks will be warmer at small radii, though the radial dependency changes with the heating mechanism. Actively heated disks typically have steeper gradients.  In either case, it is possible that the warmer inner disk would be stabilized against gravitational instability and slow down accretion. In these experiments, we find spiral arms persist in regions where the average value of $Q$ is well above that at which instability is presumed to set in, with local values exceeding this by an order of magnitude. It seem plausible that due to the global nature of the low-$m$ spiral modes, angular momentum transport may still occur in regions one would assume stable against GI. As discussed by \cite{ARS89}, $m=1$ modes can have appreciable growth rates for remarkably high values of $Q$ when the evanscent region is as much as $70\%$ of the disk radius. In the event that the GI does shut off due to increasing temperature, material from the outer, unstable portion of the disk will likely accumulate until the critical surface density for GI is reached, causing
$k_\Sigma$ to steepen. Further numerical investigation of this is necessary; we point readers to high resolution studies of disks with radial temperature gradients such as  \citet{2009SciKrum,2007ApJ...665.1254B}, \cite{2008ApJ...673.1138C}, and \cite{Krumholz2007a}.

In order  to test the effects of the gas stiffening we introduced to avoid
unphysical merging of our sink particles (see \S \ref{code_details}), we
have conducted several purely isothermal experiments in which it is
turned off.  The removal of the barotropic switch aritifically enhances accretion at early times due to sink particles formed via numerical fragmentation merging with the central star. Removing the barotropic switch is equivalent to increasing the resolution of the fragmentation process, but decreasing the resolution of the scale of fragmentation relative to $\lambda$, the disk resolution. Using a barotropic switch allows the disk to reach a higher $\lambda$ before fragmentation sets in for a given set of parameters.

\subsection{Insensitivity of disk dynamics to core temperature}\label{isothermal}

Our parameterization of disk dynamics is based on the idea that thermodynamics can be accounted for by one parameter, $\xi$, which compares the accretion rate to the disk sound speed $c_{s,d}$. A basic corollary of this notion is that the core temperature $c_{s, {\rm core}}$ has no effect on disk dynamics, except insofar as it affects the accretion rate.   We have defined $\xi$ with respect to the disk sound speed, but since
our disks and cores are the same temperature, we could equally well
have used the core sound speed. Therefore the
question arises whether $\xi$ should be computed by
normalizing the accretion rate to the disk sound speed, $c_{s,d}$, or
the core sound speed, $c_{s,c}$.  To test this, we ran simulations in which $c_{s,d}$ and $c_{s,{\rm core}}$ differed: we imposed a change in temperature over a range of radii in which infall is highly supersonic.    

To demonstrate that $\xi$ defined with respect to the disk sound speed is indeed a better predictor of the morphological and physical behavior of the
disk, we compare $\lambda$ at the time of fragmentation, $\lambda_f$ to both $\xi$ and the equivalent parameter defined in terms of the core sound speed, $\xi_{\rm core} = G \dot{M}/c_{s,{\rm core}}^3$ for runs with similar values of $\Gamma$. We observe a correlation between resolution at the time of fragmentation and $\xi$ at fixed $\Gamma$, and so if core temperature is irrelevant, these runs should follow the same trend.

\begin{figure}
    \centering
 	\includegraphics[scale=0.65]{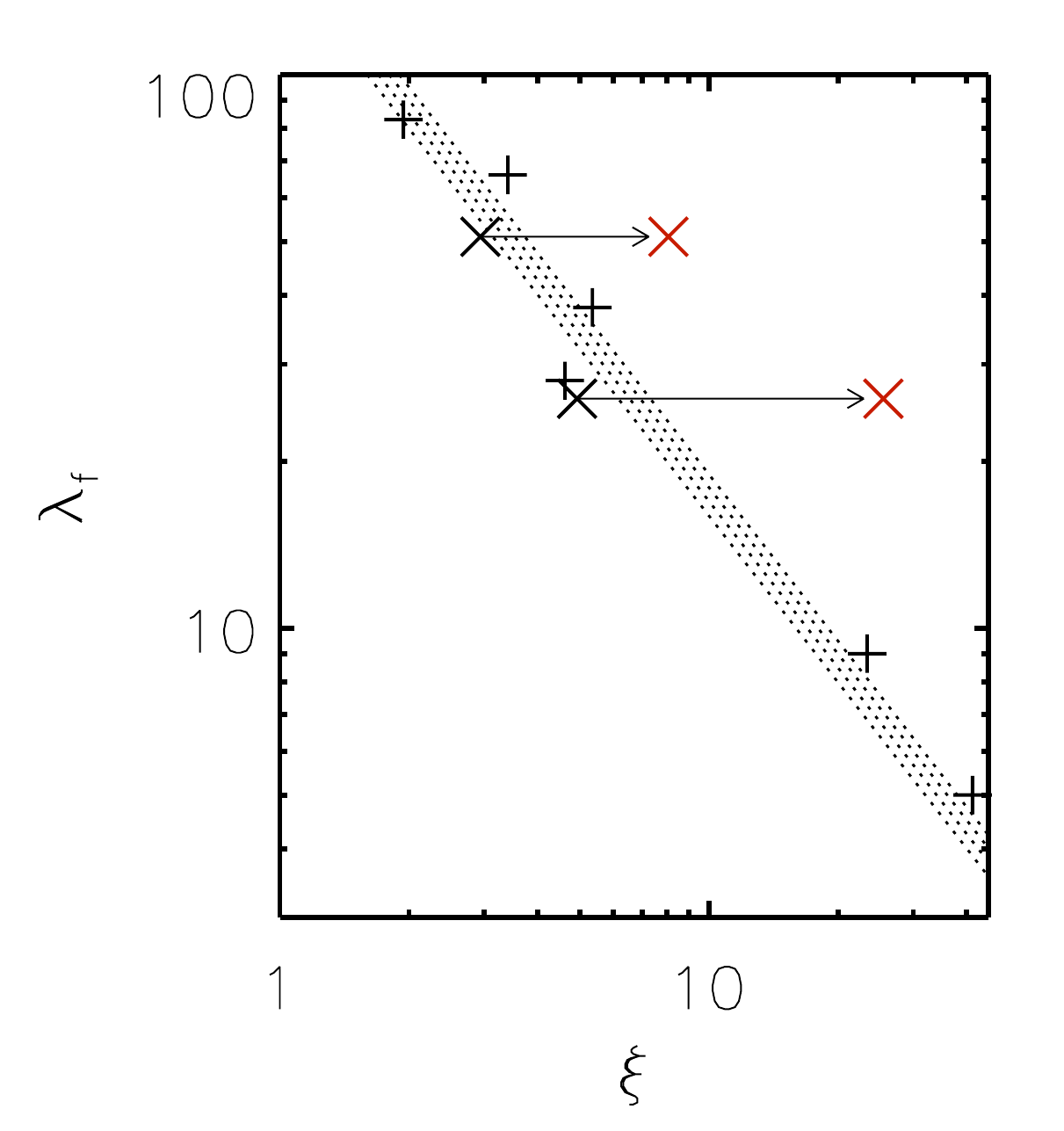}
	 
    \caption{Correlation between $\lambda_f$ and the infalling accretion rate for heated and non heated runs with comparable $\Gamma$. Plus symbols indicate non-heated runs, and the crosses are heated runs. The arrows and red crosses indicate the position of the runs evaluated with respect to $\xi_{ \rm core}$.  Runs shown have $\Gamma$ values ranging from 0.006 to 0.009. The shaded region illustrates the scaling  $\lambda_f \propto \xi^{-1}$. This scaling is related not only to the existence of a critical value of $\mu$, but also tied to the effect of resolution on fragmentation.}
        \label{heat_complot}
 \end{figure}

Figure  \ref{heat_complot}, shows that $\lambda_f$ correlates extremely well with $\xi$ at similar values of $\Gamma$, but poorly with $\xi_{\rm core}$ for the heated runs. The scaling of $\lambda_f$ with $\xi$ is also related to the existence of an upper limit on $\mu$ as a function of $\Gamma$: disks with higher $\xi$ approach this critical value of $\mu$ faster, and thus at lower $\lambda$.

 \begin{figure*}
\centering
 	\includegraphics[scale=0.65]{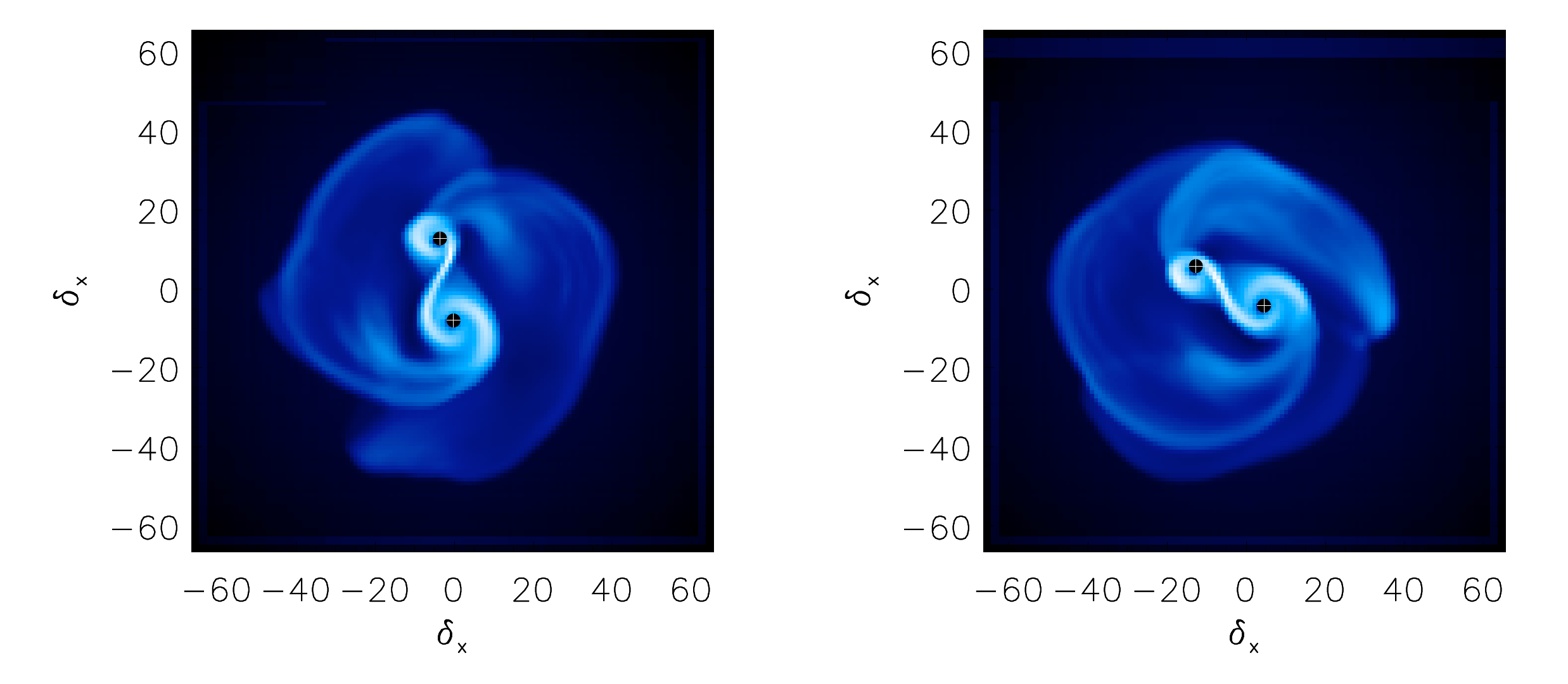}
	 
    \caption{At left: a snapshot of the standard resolution of run \#16 shortly after binary formation. At right, the same run at double the resolution. Because of the self-similar infall prescription, we show the runs at the same numerical resolution, as time and resolution are interchangeable. In this case the high-resolution run has taken twice the elapsed ``time" to reach this state. The two runs are morphologically similar and share expected disk properties. }
    \label{reso_snapshot}
 \end{figure*}

\subsection{Resolution}\label{reso}
We have shown in \S \ref{torques} that the observed accretion is consistent with the combined gravitational torques and Reynold stresses, and that these are dominant over that expected purely from numerical diffusion. Because of the self-similar infall, convergence to a steady state within a given run is a good indicator that numerics are not determining our result; in effect, every run is a resolution study. That we observe a range of behavior at the same resolution but different input parameters also implies that numerical effects are sub-dominant. We consider our disks to begin to be resolved when they reach radii such that $R_{\rm k,in}/\Delta x \geq 30$. The effective numerical diffusivity, which we plot in Figure \ref{azimalpha}, has been estimated by \cite{Krumholz04}  for ORION. Specifically they find that:
\begin{equation} \label{alpha_num}
\alpha_{\rm num} \approx 78 \frac{r_B}{\Delta x} \left({\frac{r}{\Delta x}}\right)^{-3.85}.
\end{equation} 
where
\begin{equation}
r_B = \frac{G M_* }{ c_s^2}
\end{equation}
is the standard Bondi radius.

For our typical star and disk parameters, this implies numerical $\alpha$'s of order $0.1-0.3$ at the minimum radius at which we are resolved. This implies that for our ``low" accretion rate cases, at most $1/3$ of our effective alpha could be attributed to numerical effects at low resolution.  See discussions by \cite{ Offner08,Krumholz2007a, Krumholz04} for a detailed analysis of disk resolution requirements. At our resolution of 50-100 radial cells across the disk, the dominant effect of numerical diffusion is likely a suppression of fragmentation \citep{2006ApJ...647..997S,{Nelson2006}}. Because the isothermal spiral arms can become very narrow prior to fragmentation, numerical diffusion across an arm may smear out some overdensities faster than they collapse. Therefore the conclusions regarding the fragmentation boundary are likely conservative.

We explicitly demonstrate morphology convergence in one of our binary runs. We rerun run  \#16 (as labelled in table \ref{restab}) at double the resolution ($128^3$ with 10 levels of refinement as opposed to 9). Increasing the physical resolution also decreases the code time step proportionally so that the ratio of the timestep to orbital period as a function of $\lambda$ should be preserved.  In fact, there is little that can be different between the runs at two resolutions at the same effective $\lambda$. 

The two runs have the same morphology, and characteristic disks properties as a function of $\lambda$, as expected. We show in Figure \ref{reso_snapshot} snaphots of the standard and high resolution runs. The standard resolution run (left) is at twice the elapsed ``time" of the high resolution one (right), and so the same numerical resolution, $\lambda$. We confirm that the mass accretion rate is consistent between the two runs: at the snapshots shown the mass ratio of the lower resolution run is 0.46, while the higher resolution run is 0.48. We consider this variation to be within the expected variation of the parameters (see \S \ref{selfsim_idea}). To the extent that numerical artifacts are seeding instabilities, we expect some stochasticity in the details of the fragmentation between any two runs. Although the effect is small, it is also possible that since the physical size of the disk (and the radius from which material is currently accreting)  relative to the box size is larger at the same value of $\lambda$ for the low resolution run, the large scale quadrapole potential from the image masses is stronger in the low resolution case.

We also compare a multiple run at a lower resolution by a factor of two. Again we find the same morphological outcome. We find that the disks behave equivalently at the same radial resolution although the elapsed time, and dimensional masses are different.

The scaling of $\lambda_f$ with $\xi$ in Figure \ref{heat_complot} also demonstrates that resolution plays a role in determining when disks fragment. Although the infall is self-similar, the disks approach a steady state as parameters like $Q$ and $\mu$ evolve toward constant values. This evolution, and sometimes fragmentation, is influenced by the interplay between decaying numerical viscosity and increasing gravitational instability in the disk as a function of $\lambda$. 

\section{Comparison to Previous Studies}\label{prevwork}
The literature is replete with useful simulations of protostellar and protoplanetary disks at various stages of evolution, however most involve isolated disks, without infall at large radii \citep{1994ApJ...436..335L,1996ApJ...456..279L,Rice05, LodRi05, 2006A&A...457..343F, 2006ApJ...647..997S, 2006ApJ...651..517B, 2007MNRAS.374..590L, 2008ApJ...673.1138C}.  These simulations include a wide range of physics, from magnetic fields to radiative transfer, but due to the lack of infalling matter, they neither develop disk profiles (surface density, temperature) self-consistenly, nor do they enter the regime of interest in this work: rapid accretion in the embedded phase. For a review of many of the issues addressed by current GI disk simulations, see \cite{2007prpl.conf..607D}.

There are a few simulations of self-consistent growth and evolution \citep{VorBas07, Vorobyov08}. These are ideal for following the long term evolution of more quiescient lower mass disks.  However, because they are two dimensional, and lack a moving central potential, they cannot follow the evolution of non-axisymmetric modes which are driven by the displacement of the central star from the center of mass,  nor can they accurately simulate the formation of multiple systems. Other authors have investigated the initial stage of core collapse onto disks \citep{BanPud07,1999ApJ...526..307T}, however these authors focus on the effects of magnetic fields and fragmentation of the core prior to disk formation respectively.  \cite{1999ApJ...523L.155T} and \cite{2003ApJ...595..913M} have also investigated the collapse of cores into disks and binaries, though they do not investigate many disk properties (see \S \ref{bonnorebert} for detailed comparisons). \cite{Krumholz2007a} and \cite{2009SciKrum} have conducted three dimensional radiative transfer calculations, but due to computational cost can only investigate a small number of initial conditions. 

In addition to numerical work, there are a range of semi-analytic models which follow the time evolution of accreting disks  \citep[KMK08]{2005A&A...442..703H}.  KMK08 examined the evolution of embedded, massive disks in order to predict regimes in which gravitational instability, fragmentation of the disk, and binary formation were likely. They concluded that disks around stars greater than $1-2\Msun$ were likely subject to strong gravitational instability, and that a large fraction of O and B stars might be in disk-born binary systems. \cite{2005A&A...442..703H} have also made detailed models of disk evolution, though they examine less massive disks, and do not include explicitly gravitational instability, and disk irradiation.

 In KMK08 we hypothesized that the disk fragmentation boundary could be drawn in $Q$- $\mu$ parameter space, where small scale fragmentation was characterized by low values of $Q$ and binary formation by high values of $\mu$. Due to the self-similar nature of these simulations, the distinction between these two types of fragmentation is difficult, as the continued accretion of high angular momentum material causes the newly formed fragment to preferentially accrete material and grow in mass \citep{1994MNRAS.269L..45B}. Moreover, because the disks are massive and thick, the isolation mass of fragments is comparable to the disk mass, and so there is little to limit the continued growth of fragments.

\subsection{ The evolution of the accretion parameters in the isothermal collapse of a Bonnor-Ebert Sphere}\label{bonnorebert}

 \begin{figure}
    \centering
 	\includegraphics[scale=0.4]{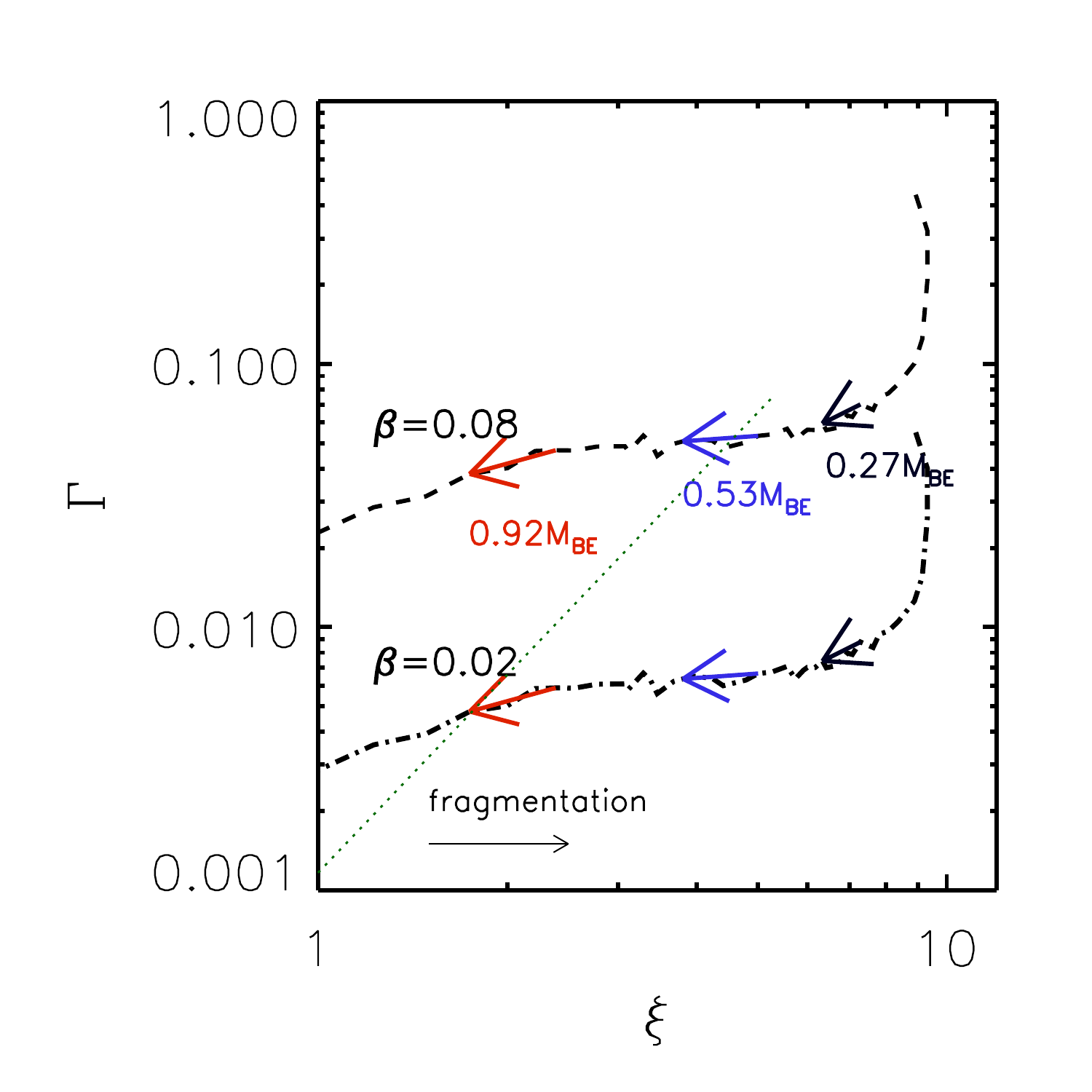}
	 
    \caption{Trajectory of a Bonnor-Ebert sphere through $\xi-\Gamma$ space.  The two lines show values of $\beta = 0.02, 0.08$ as defined in  \cite{2003ApJ...595..913M}. Arrows indicate the direction of time evolution from $t/t_{\rm ff,0} = 0-5$. $t_{\rm ff,0}$ is evaluated with respect to the central density, and arrows are labelled with the fraction of the total Bonnor-Ebert mass which has collapsed up to this point. The dotted line shows the fragmentation boundary from Figure \ref{xi_gamma}. }
    \label{bonnortrack}
 \end{figure}

While self-similar scenarios are useful for numerical experiments, they do not accurately capture the complexities of star formation. In particular, in realistic cores, $\xi$ and $\Gamma$ evolve in time. Therefore it is interesting to chart the evolution of a more realistic (though still idealized) core through our parameter space.  We consider the isothermal collapse of a Bonnor-Ebert sphere initially in solid-body rotation \citep{1956MNRAS.116..351B}. Such analysis allows us to compare our results with other numerical simulations that have considered global collapse and binary formation such as \cite{2003ApJ...595..913M} via the parameters laid out in \cite{1999ApJ...526..307T}. 

We use the collapse calculation of a 10\% overdense, non-rotating Bonnor -Ebert sphere from \cite{1993ApJ...416..303F}, and impose angular momentum on each shell to emulate solid body rotation. Figure \ref{bonnortrack} shows the trajectory of a rotating Bonnor-Ebert sphere through $\xi-\Gamma$ parameter space as a function of the freefall time $ t/t_{\rm ff,0}$, for two different rotation rates corresponding to $\beta = E_{\rm{rot}}/ E_{\rm{grav}} = 0.02, 0.08$.  The free fall time is evaluated with respect to central density.

The early spike in $\xi$ is due to the collapse of the inner flattened core at early times. Similarly, the corresponding decline in $\Gamma$ is a result of the mass enclosed increasing more rapidly than the infalling angular momentum. The long period of decreasing $\xi$ and constant $\Gamma$ arises from the balance between larger radii collapsing to contribute more angular momentum, and the slow decline of the accretion rate. This trajectory may explain several features of the fragmentation seen in  \cite{2003ApJ...595..913M}. Although not accounted for in Figure \ref{bonnortrack}, cores with high values of $\beta$ have accretion rates supressed at early times due to the excess rotational support, while those with low $\beta$ collapse at the full rate seen in \cite{1993ApJ...416..303F}.  In cores with small $\beta$,  the high value of $\xi$ may drive fragmentation while the disk is young. Alternatively, for modest values of $\beta$, $\Gamma$ may be sufficiently low while $\xi$ is declining that the disk mass surpasses the critical fragmentation threshold, and fragments via the so-called satellite formation mechanism. For very large values of $\beta$, a core which is only moderately unstable will oscillate and not collapse as seen in \cite{2003ApJ...595..913M} for $\beta > 0.3$.

\section{Discussion}
We have examined the behavior of gravitationally unstable accretion disks using three-dimensional, AMR numerical experiments with the code ORION. We characterize each experiment as a function of two dimensionless parameters, $\xi$ and $\Gamma$, which are dimensionless accretion rates comparing the infall rate to the disk sound speed and orbital period respectively. We find that these two global variables can be used to predict disk behavior, morphological outcomes, and disk-to-star accretion rates and mass ratios. In this first paper in a series we discuss the main effects of varying these parameters. Our main conclusions are:
\begin{itemize}
\item{Disks can process material falling in at up to $\xi \sim 2-3$ without fragmenting. Although increasing $\Gamma$ stabilizes disks at fixed values of $\xi$ those fed at $\xi > 3$ for many orbits tend to fragment into a multiple or binary system.}
\item{Disks can reach a statistical steady state where mass is processed through the disk at a fixed fraction of the accretion rate onto the disk. The discrepancy between these two rates, $\mu$, scales with $\Gamma$; disks with larger values of $\Gamma$ can sustain larger maximum disk masses before becoming unstable. The highest disk mass reached in a non-fragmenting system is $\mu \approx 0.55$ or $M_* \sim M_d$.}
\item{Gravitational torques can easily produce effective accretion rates consistent with a time averaged $\alpha \approx 1$.}
\item{The minimum value  of $Q$ at which disks begin to fragment is roughly inversely proportional to the disk scale height. It is therefore important to consider not only $Q$ but another dynamical parameter when predicting fragmentation, at least in disks which are not thin and dominated by axisymmetric modes.}
\item{The general disk morphology and multiplicity is consistent between isothermal runs and irradiated disks with similar effective values  of $\xi$.}
\end{itemize}

These conclusions are subject to the qualification that fragmentation occurs for lower values of $\xi$ as the disk resolution increases, and so it is possible that the location of the fragmentation boundary will shift with increasing resolution. However we expect that our results are representative of real disks and other numerical simulations in so far as they have comparable dynamic range of the parameters relevant to fragmentation such as $\lambda_J/ \lambda$. 

\section{Acknowledgments}
The authors would like to thank Chris McKee, Jonathan Dursi, Stella Offner, Andrew Cunningham, Norman Murray, and Yanqin Wu for insightful discussions and technical assistance. KMK was funded in part by a U. of T. fellowship. CDM received support through an Ontario Early Research Award, and by NSERC Canada. MRK received support for this work from an Alfred P. Sloan Fellowship, from NASA, as part of the Spitzer Theoretical
Research Program, through a contract issued by the JPL, and from the National Science Foundation, through grant AST-0807739. RIK received support for this work provided by the US Department of Energy at Lawrence Livermore National Laboratory under contract B-542762; NASA through ATFP grants NAG 05-12042 and NNG 06-GH96G and NSF through grant AST-0606831. All computations were performed on the Canadian Institute for Theoretical Astrophysics Sunnyvale cluster, which is funded by the Canada Foundation for Innovation, the Ontario Innovation Trust, and the Ontario Research Fund.  This research was supported in part by the National Science Foundation under Grant No. PHY05-51164.

\end{document}